\documentclass[prl,aps,twocolumn, superscriptaddress]{revtex4-2}
\usepackage{amsfonts}
\usepackage{amssymb}
\usepackage{amsmath}
\usepackage{graphicx}
\usepackage{verbatim}
\usepackage{physics,bm}
\usepackage{lipsum, wrapfig}
\usepackage{enumitem}
\usepackage{hyperref}
\usepackage[dvipsnames]{xcolor}
\usepackage{soul}
\usepackage[normalem]{ulem} 
\usepackage{textcomp}

\usepackage{bibunits}
\defaultbibliographystyle{apsrev4-2}
\defaultbibliography{Z3Driven}
\makeatletter
\let\rtx@bibliography\@gobble
\let\bibliography\@gobble
\makeatother

\usepackage{color}
\usepackage{xcolor,colortbl}
\usepackage{comment}
\newcommand {\nc} {\newcommand}
\nc {\ER} [1]{{\color{orange}\footnotesize\textsc{(er)} #1} }

\hypersetup{pdflang={English},colorlinks=true,linkcolor=Cerulean,citecolor=RoyalBlue,urlcolor=RoyalBlue,breaklinks=true}

\usepackage[titles]{tocloft}
\usepackage{titlesec}

\begin{document}

\makeatletter
\let\origaddcontentsline\addcontentsline
\renewcommand{\addcontentsline}[3]{}  
\makeatother

\title{A Qutrit Time Crystal Stabilized with Native Chiral Interactions}
\author{Noah Goss}
\thanks{These authors contributed equally to this work. Correspondence should be addressed to \texttt{noahgoss@berkeley.edu} and \texttt{nsuri@lbl.gov}}
\affiliation{Department of Physics, University of California, Berkeley, Berkeley, CA 94720, USA.}
\affiliation{Computational Research Division, Lawrence Berkeley National Laboratory, Berkeley, CA 94720, USA.}

\author{Nishchay Suri}
\thanks{These authors contributed equally to this work. Correspondence should be addressed to \texttt{noahgoss@berkeley.edu} and \texttt{nsuri@lbl.gov}}
\affiliation{Computational Research Division, Lawrence Berkeley National Laboratory, Berkeley, CA 94720, USA.}

\author{Brian Marinelli}
\affiliation{Department of Physics, University of California, Berkeley, Berkeley, CA 94720, USA.}
\affiliation{Computational Research Division, Lawrence Berkeley National Laboratory, Berkeley, CA 94720, USA.}

\author{Larry Chen}
\affiliation{Department of Physics, University of California, Berkeley, Berkeley, CA 94720, USA.}
\affiliation{Computational Research Division, Lawrence Berkeley National Laboratory, Berkeley, CA 94720, USA.}

\author{Akel Hashim}
\affiliation{Computational Research Division, Lawrence Berkeley National Laboratory, Berkeley, CA 94720, USA.}

\author{Sajant Anand}
\affiliation{Department of Physics, Harvard University, Cambridge, MA 02138, USA.}

\author{Alexis Morvan}
\affiliation{Google Quantum AI, Santa Barbara, CA 93117, USA}

\author{Ravi K. Naik}
\affiliation{Computational Research Division, Lawrence Berkeley National Laboratory, Berkeley, CA 94720, USA.}

\author{Ermal Rrapaj}
\affiliation{Computational Research Division, Lawrence Berkeley National Laboratory, Berkeley, CA 94720, USA.}

\author{David I. Santiago}
\affiliation{Computational Research Division, Lawrence Berkeley National Laboratory, Berkeley, CA 94720, USA.}

\author{Wibe de Jong}
\affiliation{Computational Research Division, Lawrence Berkeley National Laboratory, Berkeley, CA 94720, USA.}

\author{Norman Y. Yao}
\affiliation{Department of Physics, Harvard University, Cambridge, MA 02138, USA.}

\author{Joel E. Moore}
\affiliation{Department of Physics, University of California, Berkeley, Berkeley, CA 94720, USA.}
\affiliation{Materials Sciences Division, Lawrence Berkeley National Laboratory, Berkeley, CA 94720, USA.}

\author{Irfan Siddiqi}
\affiliation{Department of Physics, University of California, Berkeley, Berkeley, CA 94720, USA.}
\affiliation{Computational Research Division, Lawrence Berkeley National Laboratory, Berkeley, CA 94720, USA.}
\begin{bibunit}

\begin{abstract}

Periodically driven quantum many-body systems can spontaneously break discrete time-translation symmetry, realizing discrete time crystals. 
To date, both experimental and theoretical efforts have largely focused on the simplest case of spontaneous period-doubling in $\mathbb{Z}_2$ discrete time crystals realized with qubits.
This owes, in part, to the challenge of stabilizing eigenstate order in higher discrete symmetry ($\mathbb{Z}_n$) time crystals, due to the presence of richer domain wall physics. 
Here, we demonstrate the realization of a $\mathbb{Z}_3$ discrete time crystal by implementing a Floquet chiral clock model in a chain of 15 superconducting qutrits. 
Unlike the conventional Ising setting, our system features a tunable chiral angle that governs domain-wall dynamics, spectral degeneracies, and crucially, the stability of time-crystalline order. 
Using disordered nearest-neighbor chiral interactions, we observe robust subharmonic period tripling that persists across a wide range of drive strengths and is independent of initial state. 
Finally, we highlight the special role that chirality plays in our $\mathbb{Z}_3$ discrete time crystal---in its absence, the system's Floquet dynamics exhibit a marked initial state dependence governed by domain wall degeneracies. 
Our results establish native qudit hardware as a powerful platform to access a broader landscape of non-equilibrium phases.
\end{abstract}

\maketitle

\begin{figure*}[t!]
    \centering
    \includegraphics[angle = 0, width=1.0\linewidth]{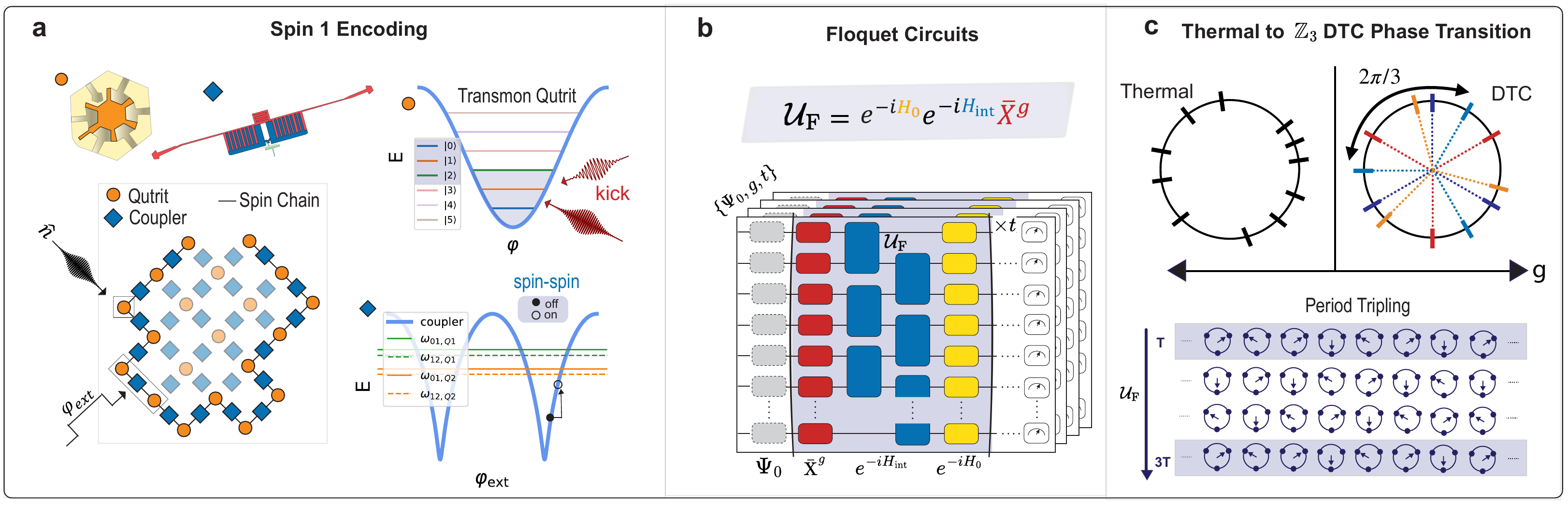}
    \caption{\textbf{Experimental concepts.} \textbf{a}, We engineer a spin-1 chain on a superconducting quantum processor composed of 20 flux-tunable transmons (orange) and tunable couplers (blue). Local ``kicking" (red) is performed by charge driving ($\hat{n})$ coherent Rabi oscillations at the $\ket{0}\leftrightarrow\ket{1}$ and $\ket{1} \leftrightarrow \ket{2}$ transition frequencies. The disordered spin-spin interaction Hamiltonian $H_{\text{int}}$ (light-blue) is realized from fast flux pulses  on the tunable couplers ($\varphi_{\text{ext}}$) which turns on a large dispersive coupling between the qutrits. \textbf{b}, A schematic illustration of the experimental circuits used to implement the Floquet $\mathbb{Z}_3$ CCM. In addition to the kicking and spin-spin interaction described in the previous panel, we also engineer disordered on-site fields ($H_0$) via virtual frame updates to our microwave pulses. To study the eigenstate ordering of our system across the thermal-DTC phase transition, we vary the choice of initial trit-string state $\Psi_0$, kicking strength $g$, and time steps applying the Floquet unitary $\mathcal{U}_F$. $\textbf{c}$, (top) While in the thermal phase the quasienergies are not ordered, in the $\mathbb{Z}_3$-DTC phase the entire eigenspectrum is paired off into triplets, where each quasienergy is related to its cousins by $\pm 2\pi/3$. (bottom) The characteristic time domain response in the $\mathbb{Z}_3$-DTC phase is defined by subharmonic period tripling, where all initial spin sites (depicted as clocks) return to their original configuration after 3 applications of $\mathcal{U}_F$. }
    \label{fig:fig1}
\end{figure*}

\section*{Introduction}

Out-of-equilibrium many-body physics has emerged as a fertile arena for discovering phenomena with no equilibrium counterpart.
The simplest approach to taking a system out of equilibrium is to periodically drive it. Even in this Floquet setting, new phenomena can emerge such as novel symmetry protected topological phases~\cite{harper2020topology}, Floquet topological insulators~\cite{rudner2020band}, and spontaneously broken time-translation symmetry giving rise to discrete time crystals (DTCs) ~\cite{PhysRevLett.116.250401, else2016floquet, yao2017discrete,zaletel2023colloquium,else2020discrete,PhysRevB.94.085112}. This latter phenomena has attracted a tremendous amount of theoretical and experimental attention, especially in the context of $\mathbb{Z}_2$ discrete time crystals, where period doubling naturally emerges in many-body Floquet systems realized with qubits~\cite{mi2022time, frey2022realization, google_edge,xiang2024long,zhang2022digital,jin2025topological,will2025probing,zhang2017observation,choi2017observation,PhysRevLett.120.180602,PhysRevLett.120.180603,he2025experimental, kyprianidis2021observation}. While nascent experiments have explored the possibility of higher discrete symmetry $\mathbb{Z}_n$ DTCs~\cite{choi2017observation}, robustly stabilizing these phases can be subtle. In particular, in a $\mathbb{Z}_n$ DTC, domain walls exhibit a natural handedness that, without the breaking of certain symmetries, can introduce degeneracies in the quasienergy spectrum, preventing DTC stability~\cite{surace2019floquet, Bennett}. 

In this work we go beyond the Ising paradigm to explore the stabilization of a $\mathbb{Z}_3$ DTC in a chiral clock model (CCM). Such models feature chiral interactions which introduce directionality and frustration in domain wall dynamics~\cite{ Fendley_2012,Fendley_stability}, support incommensurate phases~\cite{howes,PhysRevLett.49.793,PhysRevB.24.398, Norm_Parafermion, zhang2025probing}, and host critical behavior outside the Ising universality class~\cite{Sachdev_z3,PhysRevB.98.205118,Parameswaran_localization, keesling2019quantum}.
Despite these remarkable properties, experimental realizations of CCMs have remained largely elusive due to the challenges associated with their physical implementation. To address this, we employ multi-level superconducting circuits~\cite{scrambling, brock_quantum_2025, goss2022high, champion_multifrequency_2025, ticea2026} -- or qudits --  which enable direct realizations of higher spin models without the added connectivity and compilation overhead associated with qubit-based hardware~\cite{Sawaya2020, PhysRevA.109.012426, camacho2024observing}.

With our native $\mathbb{Z}_3$ CCM, we directly investigate the essential nature of chirality in Floquet dynamics. In driven Ising models, disordered interactions are necessary to induce many-body localization (MBL)~\cite{pal,ponte,PhysRevLett.113.107204}, which can stabilize a DTC. Surprisingly, in our driven CCM, disordered couplings alone are not sufficient to stabilize a $\mathbb{Z}_3$ DTC due to richer domain wall physics governed by the chiral interactions. With our qutrit processor, we precisely engineer both the chirality and coupling disorder, enabling the observation of a $\mathbb{Z}_3$ DTC phase, whose lifetime (in the absence of decoherence) is expected to scale exponentially with system size. 
In the presence of chiral couplings, we observe robust period tripling that persists across a wide range of drive strengths and is independent of initial state, characteristic of so-called eigenstate order. 
By contrast, when we remove chirality, we introduce degeneracies between two distinct antiferromagnetic domain walls that break the $\mathbb{Z}_3$ DTC stability, which we directly observe through initial state dependence.
In addition, we investigate the continuous interpolation between a $\mathbb{Z}_2$ and $\mathbb{Z}_3$ DTC within the qutrit state space. 
Our results establish qudit platforms as a compelling setting for exploring non-equilibrium phases beyond $\mathbb{Z}_2$, and point toward a broader landscape of dynamical orders within reach of near-term quantum simulators.

\section*{Results}

\subsection{Model Hamiltonian and Experimental Realization} 
We realize a spin-1 driven chiral clock model on a chain of $N$ superconducting qutrits (Fig.~\ref{fig:fig1}a), where the spin $\ket{s} \in \{\ket{0},\ket{1},\ket{2}\}$ is encoded in the three lowest transmon levels. Our device consists of a square grid of 20 flux-tunable transmon qutrits \cite{abdurakhimov2024technologyperformancebenchmarksiqms}. Microwave drives and virtual frame updates are employed to implement the unitaries representing, respectively, the global kick $\bar{X}^g$ and the onsite fields $e^{-i H_0}$
\begin{align}
    \bar{X}^g &= \prod_j X_j^g \;, &  H_0  &= \sum_j {h_{j} e^{i \varphi_j} Z_j + \text{h.c.}} \;,
\end{align} 
composed of the qutrit clock $Z$ and shift $X$ operators, where $Z\ket{s} = \omega_3^s\ket{s}$, $X\ket{s} = \ket{s \oplus_3 1}$, and $\omega_3=e^{2\pi i/3}$ is the cube root of unity. Nearest neighbor interactions are mediated by flux-tunable couplers \cite{PRXQuantum.4.010314}. Fast flux pulsing of the couplers activates a strong dispersive interaction \cite{PhysRevLett.130.030603} and produces a native chiral clock Hamiltonian
\begin{align}\label{eq:hint}
    H_{\text{int}}&=\sum_{j} \underbrace{J_je^{i \theta_j} Z_j Z_{j+1}^\dagger}_{\mathbb{Z}_3  \text{ preserving}}+ {J_j' e^{i \theta_j '} Z_j Z_{j+1}} + \text{h.c.} \;
\end{align}
(see Supplementary Materials for more details). This spin-spin interaction Hamiltonian admits two terms, the first of which is the typical chiral clock interaction. This term preserves the $\mathbb{Z}_3$ symmetry by commuting with $\bar{X} = \prod_j X_j$ and generates unique phenomena in CCMs, motivating decades of theoretical research, especially in the static and uniform coupling limit~\cite{Fendley_2012,Fendley_stability,howes,PhysRevLett.49.793,PhysRevB.24.398,PhysRevB.98.205118}. Our system Hamiltonian also admits a second term that breaks this symmetry, and has, to the best of our knowledge, not appeared before in the literature. 
Since $Z$ is non-Hermitian for systems with more than two levels, the couplings acquire a directional, or \emph{chiral},  phase $\theta_j,\theta^\prime_j$. Combining all of these operations, we realize a short-range driven chiral clock model by implementing the time-periodic Floquet unitary $\mathcal{U}_F$ 
\begin{align}
    \mathcal{U}_F & = e^{-iH_{0}} e^{-iH_{\text{int}}}\bar{X}^g \;,
\end{align}
as shown in Fig.~\ref{fig:fig1}b.

\begin{figure*}[t!]
    \centering
    \includegraphics[width=1\linewidth]{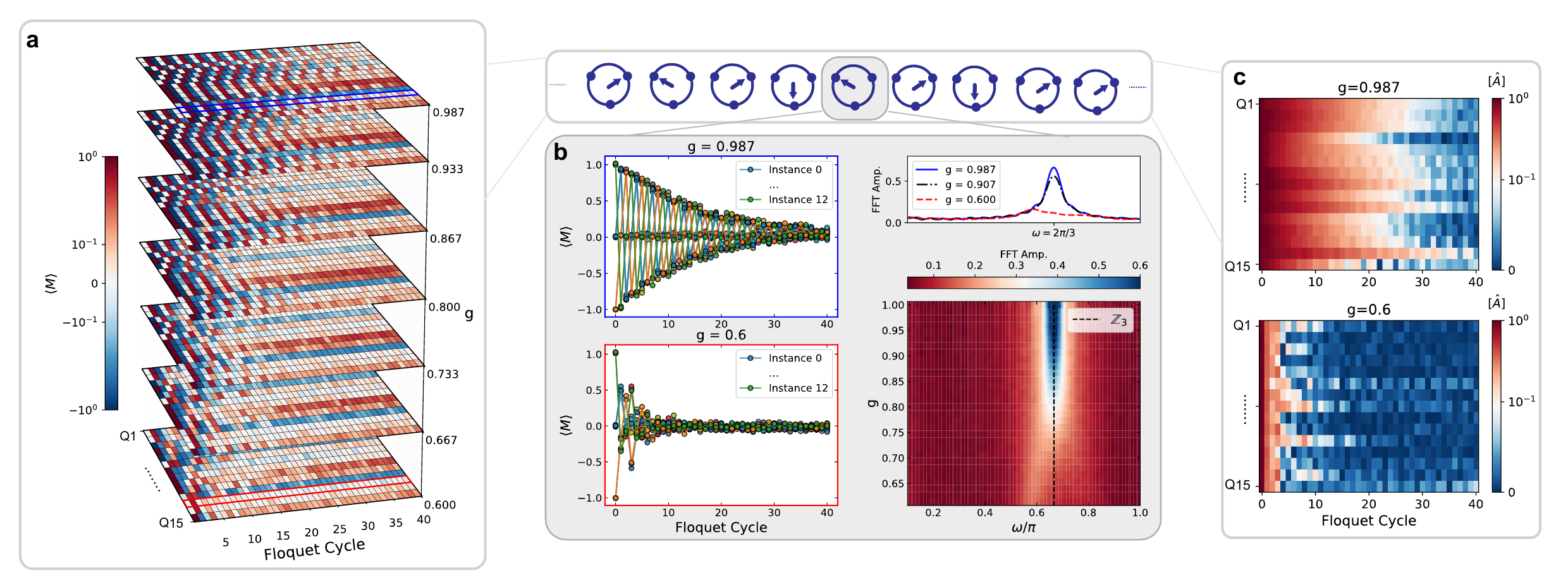}
    \caption{\textbf{$\mathbf{Z}_3$ DTC in a 15 qutrit chain.}  \textbf{a}, Spin-$1$ magnetization of a chain of $15$ qutrits as a function of Floquet cycle for multiple values of kicking strength $g$. \textbf{b}, (left) Single qutrit marginalized spin-$1$ magnetization of Q$13$ in the DTC-phase (blue) and thermal phase (red) where instance refers to different prepared initial states. (right) Fourier transform of the spin-1 magnetization as a function of kicking strength $g$. (top) Change in the Fourier peak as the kicking strength $g$ is varied. \textbf{c}, Instance-averaged autocorrelator $[\hat A(t)]$, where $\hat{A}(t)=\omega_3^{-t} \langle Z(t) Z^\dagger(0)\rangle$ for the entire chain in the DTC and thermal phases.} 
    \label{fig:fig2}
\end{figure*}
\subsection*{$\mathbb{Z}_3$ Discrete Time Crystal Eigenstate Order}

At $g=1$, the drive $\bar{X}^g$ performs a perfect global kick, rotating all spins by $2\pi/3$ to the next clock state and perfectly dynamically-decoupling the $\mathbb{Z}_3$ non-preserving terms in Eq.~\ref{eq:hint}.
As shown in Fig.~\ref{fig:fig1}c, after three kicks, the spins return to their initial configuration, yielding a subharmonic response with period tripling. The entire quasienergy spectrum of $\mathcal{U}_F$ is ordered into triplets (see Supplementary Materials for derivation) $\{ \varepsilon, \varepsilon+2\pi/3, \varepsilon - 2\pi/3\}$ with each quasienergy~\cite{surace2019floquet}
\begin{align}
    \varepsilon = 2\sum_j J_j \cos(2\pi(s_{j}-s_{j+1})/3+\theta_j) \;.
    \label{eq:4}
\end{align} 
Every such eigenstate is a long-range ordered cat state. 
Crucially, away from the exactly soluble point, $g = 1$, the Floquet dynamics no longer exhibit a $\mathbb{Z}_3$ spin-rotation symmetry and are determined by all of the terms in $\mathcal{U}_{F}$~\cite{zaletel2023colloquium,else2016floquet}.
However, the $\mathbb{Z}_3$ non-preserving terms are still approximately decoupled when $g \sim 1$; thus,  while the $(J_j',\theta_j',h_j, \varphi_j)$ terms are always present, the Floquet dynamics are dominated by the couplings $(J_j, \theta_j)$, which we explicitly focus on throughout the text.

Demonstrating true time-crystalline behavior requires robust period tripling extending beyond the trivial $g=1$ point, where deviations in $g$ act as a local perturbation. 
The presence of strongly disordered couplings pushes the system toward the many-body localized (MBL) regime,
both helping to prevent Floquet heating effects and ensuring that local perturbations do not destroy the cat-state nature of the eigenstates~\cite{pal,ponte}.
Additionally, as will be discussed in detail in the following section, avoiding degeneracy in the Floquet quasienergies requires the chiral angles $\theta_j$ to be detuned away from integer multiples of $\pi/3$. With such criteria met, even away from $g=1$, the quasienergy triplet pairing and long-range cat eigenstates can remain stable up to exponentially small corrections with system size $N$ (see Supplementary Information for additional details), satisfying the criteria for a stable discrete time-crystalline phase.

\begin{figure*}
    \centering
    \includegraphics[width=\linewidth]{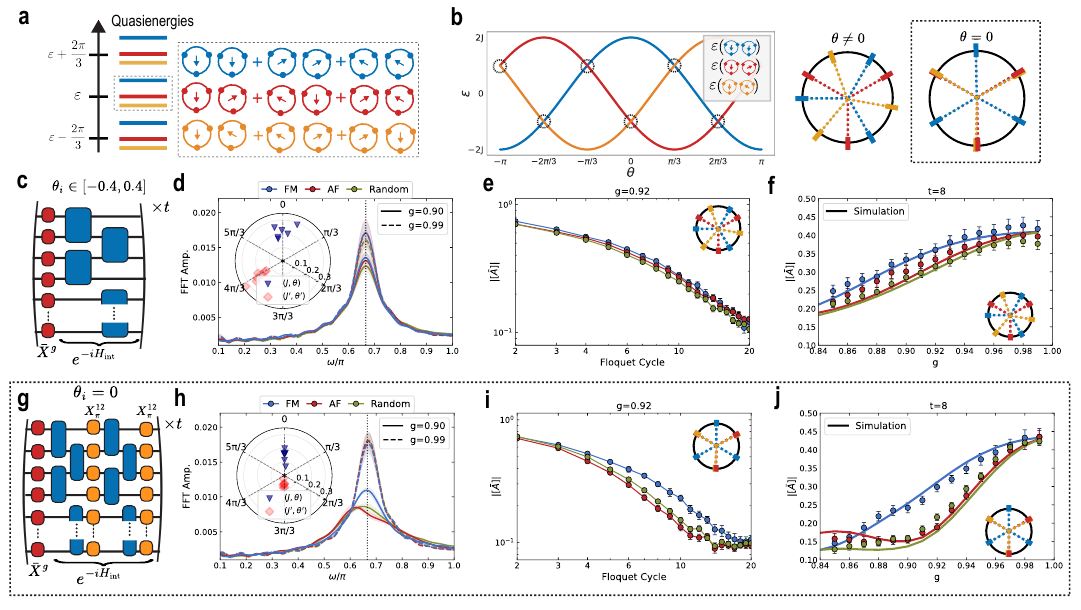}
    \caption{\textbf{Demonstrating the importance of chirality in DTC stability.} \textbf{a}, Quasienergy ($\epsilon$) spectrum and corresponding eigenstates of two clock qutrits. The FM eigenstate is shown with blue clock spins, and the two AF states are shown with the red and orange clock spins. \textbf{b}, Quasienergies for the displayed eigenstates as a function of chiral angle $\theta$ with regions of degeneracy circled. \textbf{c}, Circuit implementation of $\mathcal{U}_F$ for $e^{-i H_{\text{int}}}$ with disordered $\theta_j$. \textbf{d}, Site and initial state averaged $\expval{M}$ FFT for FM, AF, and randomly prepared tritstring states with the experimental couplings corresponding to $H_{\text{int}}$ displayed on an inset radial plot for an $8$ qutrit chain. \textbf{e}, Site and initial state averaged autocorrelator $|[\bar{A}]|$ response as a function of time for a kick of $g=0.92$. \textbf{f}, $\|[\bar{A}]|$ response at $t=8$ as a function of kicking strength $g$. Noiseless numerical simulations are globally rescaled to account for system decoherence. \textbf{g} Circuit implementation for $\mathcal{U}_F$ where $e^{-i H_{\text{int}}}$ is modified to set $\theta_j \rightarrow 0$ via local dynamical decoupling pulses (orange). The spin-spin gates (blue) are split in half with the DD pulses symmetrizing the interaction. \textbf{h}-\textbf{j}, the same analysis as \textbf{d}-\textbf{f}, with chiral couplings turned off. The stable nature of the FM states relative to AF and randomly prepared trit string states can be observed across the FFT and autocorrelator responses of the sites. }
    \label{fig:fig3}
\end{figure*}

Leveraging our native control over these critical elements of the $\mathbb{Z}_3$-DTC, we realize this phase of matter in a chain of 15 superconducting qutrits as displayed in Fig.~\ref{fig:fig2}. We engineer disorder by choosing flux pulse parameters that randomly select spin-spin couplings with $J_j \in [0.08,0.25]$ and $\theta_j \text{ mod }\pi/3 \in [0.125, 0.9]$. We demonstrate a continuous tuning of our system from a thermal phase to the $\mathbb{Z}_3$-DTC phase In Fig.~\ref{fig:fig2}a, for a random initial trit-string state, we display the full chain spin-1 magnetization $M = \ketbra{0} - \ketbra{2}$ response (which, unlike the clock $Z$, is hermitian and therefore observable) across our qutrit chain as the kicking strength is continuously tuned away from $g=1$. In total, we prepare 13 different initial trit string states, including the 3 ferromagnetic states $\ket{\Psi}_j = \ket{j}^{\otimes n}, j\in \mathbb{Z}_3$ and 10 randomly chosen $n$-qutrit trit string states. Our deepest circuits (up to 40 Floquet cycles) correspond to 80 cycles of entangling gates, utilizing 600 total two-qutrit gates. Remarkably, as exhibited in Fig.~\ref{fig:fig2}b, the robust nature of this phase of matter can be observed in the fast Fourier transform (FFT) of the $M$ response to $\mathcal{U}_F$ at an individual bulk site, where the magnetization responds robustly at $\omega = 2\pi/3$ even as the kicking strength $g$ is tuned significantly away from the special point ($g=1$) until eventually melting into the thermal phase. The characteristic time domain response of the two phases can be observed in the left side of Fig.~\ref{fig:fig2}b, where in the $\mathbb{Z}_3$-DTC phase, robust period tripling is observed for all initial states, whereas in the thermal phase all initial states rapidly thermalize. Finally, as depicted in the autocorrelator expectation values in Fig.~\ref{fig:fig2}c, this continuous tuning between a thermal and $\mathbb{Z}_3$-DTC phase can be observed across all bulk-spin sites, further confirming the lock-step behavior of bulk sites up to expected fluctuations in the individual transmon lifetimes. To certify the stability of our DTC behavior, we employ dynamical decoupling to invert the time evolution from our two-qutrit gates \cite{PhysRevLett.134.050601}, and confirm that the dampening of oscillations in our DTC is the product of intrinsic system decoherence (see the Supplementary Materials for more details).

\subsection*{Chirality Determined Stability of Time Crystal}

A crucial component for the stability of time crystals is that the eigenstates of $\mathcal{U}_F$ are long range ordered cat states~\cite{else2016floquet}.
If a degeneracy is introduced into the Floquet spectrum, these cat states can hybridize to a short-range correlated eigenstate, which in turn will not show stable time crystalline behavior.

\begin{figure}
    \centering
    \includegraphics[width=0.9\linewidth]{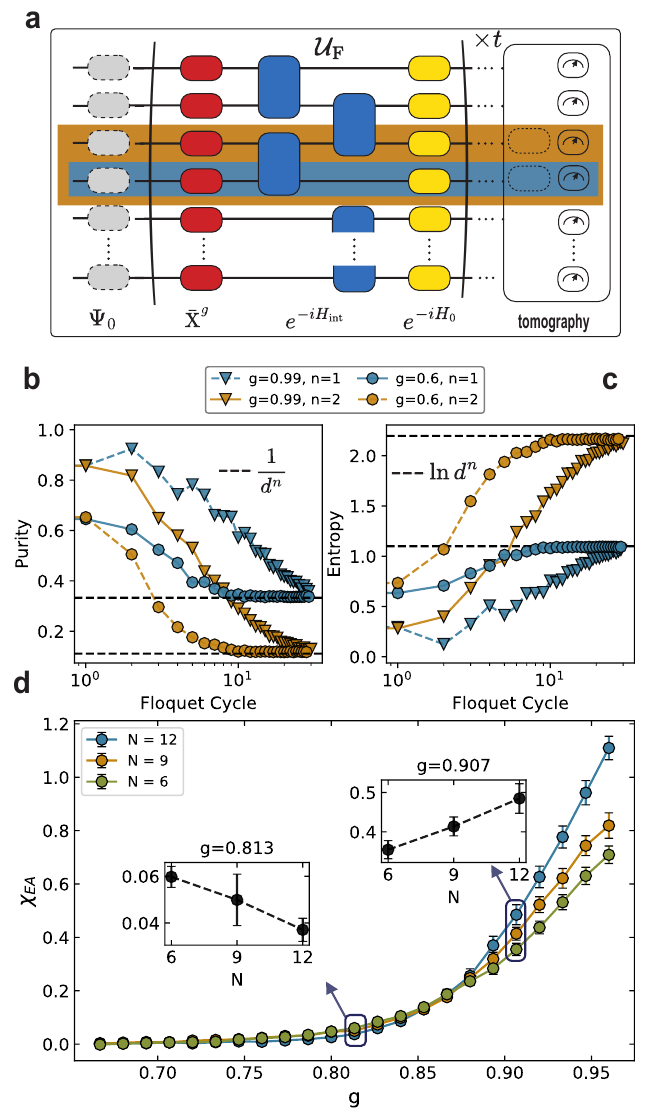}
    \caption{\textbf{System level probes of thermal-$\mathbb{Z}_3$ DTC phase transition.} \textbf{a}, Schematic of Floquet circuits $\mathcal{U}_F$ where at each Floquet cycle $t$, the bulk spin site response is probed with 1 or 2 qutrit state tomography. \textbf{b,c}  For a subset of $n=1$ and $n=2$ qutrits from the chain, respectively the tomographically reconstructed density matrix state purity $(\text{Tr}[\rho^2])$ and entropy ($-\text{Tr}[\rho\log\rho]$) response as a function of Floquet cycle in both the $\mathbb{Z}_3$-DTC ($g=0.99$) and thermal ($g=0.6$) regimes. Dashed horizontal lines depict the expected values for maximally mixed states $\rho = \frac{1}{d^n}\text{I}_{d^n}$ \textbf{d}, Spin glass order parameter ($\chi_{\text{EA}}$) as a function of kicking strength $g$ and qutrit chain length $N$. We observe a crossing in the range of $0.82 \lesssim g_c \lesssim 0.89$, similar to our numerical modeling of the system. The data points and error bars represent respectively the mean and standard error of $\chi_{EA}$ calculated across all 13 instances of initial trit-string state ($\Psi_0$).}
    \label{fig:fig4}
\end{figure}

In our Floquet model, the chiral angle couplings $\theta_j$ determine stability by imposing different energy costs on domain walls with different magnetic order. This can be observed by analyzing a system of two interacting clock spins ($s_1$,$s_2$).  
There exist three different types of domain walls corresponding to spin differences $s_2-s_1 \in \{0,1,-1\}$, where $0$ represents ferromagnetic (FM) ordering and $\pm 1$ represent different anti-ferromagnetic (AF) orderings. We plot their quasienergies at $g=1$ (see Eq.~\ref{eq:4}) as a function of chiral angle $\theta$ in Fig~\ref{fig:fig3}(b).
At $\theta=0$, the two AF domains become degenerate and are energetically distinct from the FM domain. As a result, for uniform $\theta_j =0$, two eigenstates with different flavors of AF domains become equal in energy and only the FM states remain energetically distinct. We therefore expect that all non-FM ordered states will experience short-range hybridization leading to unstable DTC behavior.

We highlight this special role of chirality in determining the $\mathbb{Z}_3$ DTC stability by experimentally comparing the disordered $\theta_j \neq 0$ (Fig.~\ref{fig:fig3}c-f) with the uniform $\theta_j= 0$ case (Fig.~\ref{fig:fig3}g-j) in an 8 qutrit chain. To probe the contrasting behavior, we analyze the Floquet dynamics of 3 FM, 6 AF, and 10 random initial trit string states. 
For the $\theta_j \neq 0$ case, we implement $\mathcal{U}_F$ as shown in Fig.~\ref{fig:fig3}c, with the parameters plotted in panel d. 
With these chiral couplings, we observe a robust period tripling response for a wide range of kicking strengths $g$ across all initial states, confirmed by the chain averaged magnetization $\expval{M}$ spectroscopy in Fig.~\ref{fig:fig3}d. We further confirm that the site and initial state averaged autocorrelator $|[\bar{A}]|$ (Fig.~\ref{fig:fig3}e-f) shows minimal initial state dependence across a range of Floquet cycles and kicking strengths.
By contrast, when we remove chirality by applying local DD pulses to set $\theta_j \rightarrow 0$ across the chain (Fig.~\ref{fig:fig3}g), we directly observe the preferred stability of the FM ordered states compared to AF and random trit string states (Fig.~\ref{fig:fig3}h-j). In Fig.~\ref{fig:fig3}h, we observe that at $g=0.9$, only the FM ordered states respond with period tripling.
Moreover, this difference in stability between the FM states and different magnetic orders is observable in $|[\bar{A}]|$ even at early time steps (Fig.~\ref{fig:fig3}i) and persists across a wide range of kicking strengths (Fig.~\ref{fig:fig3}j), in close agreement with numerical simulation. This dependence on chirality to determine DTC stability highlights the richness of qudit clock model physics with no analogous counterpart in widely studied $\mathbb{Z}_2$ Ising models.

\subsection*{System Probes of Phase Transition}

\begin{figure}[t!]
\centering
    \includegraphics[width=1\linewidth]{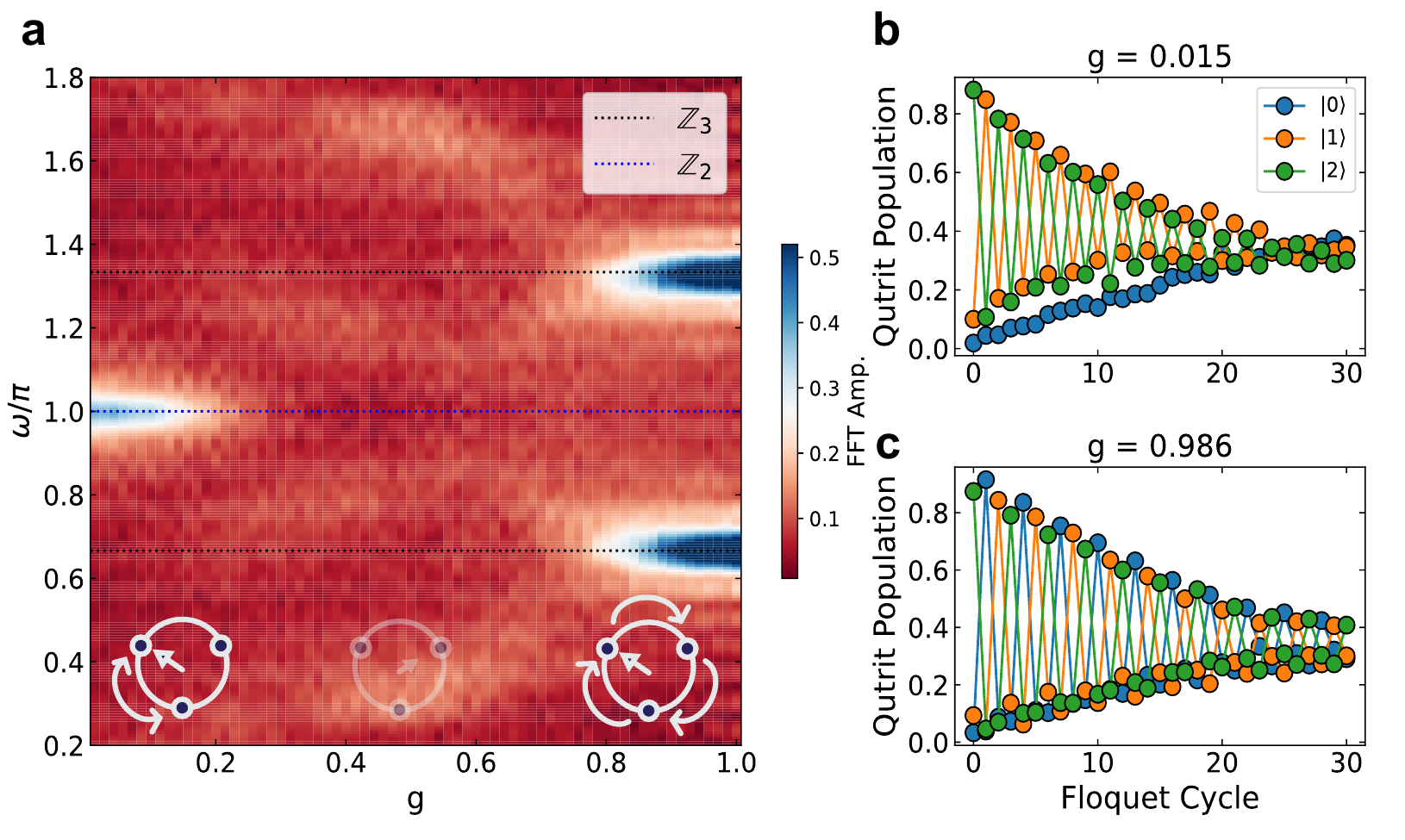}
    \caption{\textbf{Tuning from a $\mathbb{Z}_2$ to $\mathbb{Z}_3$ DTC.} \textbf{a}, The spin-1 magnetization vs. time FFT of one spin site in an 8 qutrit chain where our system is now driven with the modified global kicking $\bar{X}^g = \bar X_{12}^{1-g} \bar X^g$. As can be observed from the FFT, for a small $g$, the system responds as a $\mathbb{Z}_2$ DTC with characteristic period doubling, before thermalizing at intermediate $g$, and finally recovering as a $\mathbb{Z}_3$ DTC at large $g$. \textbf{b}-\textbf{c} The qutrit state populations vs. Floquet cycle in respectively the $\mathbb{Z}_2$ and $\mathbb{Z}_3$ DTC regions. }
    \label{fig:fig5}
\end{figure}

Dynamical phase transitions in non-equilibrium many-body systems are characterized by stark differences in the dynamics of the system and are typically defined in the thermodynamic limit as the chain length $N \rightarrow \infty$. To probe the nature of these phases in an experimental setting, we employ qutrit tomography to study entanglement growth and investigate the finite system scaling behavior to confirm a genuine phase transition.

State tomography is a powerful tool for probing and contrasting the bulk entanglement growth in the thermal and the DTC phase. The MBL nature of the DTC phase prevents ergodicity, giving rise to extensive local integrals of motion that are dressed localized clock operators~\cite{PhysRevB.90.174202}. In the DTC phase, the interaction between these dressed operators falls off exponentially with their separation, leading to a slow growth of entanglement ($ \sim \log(t)$). By contrast, these operators delocalize in the thermal case causing a fast entanglement growth ($\sim t$).

To demonstrate this contrast in the behavior of the phases, we perform full one and two qutrit state tomography of bulk sites for all times $t$ (Fig.~\ref{fig:fig4}a). Despite the intrinsic decoherence in the system, the pronounced nature of the difference in entanglement growth between the two phases is still clearly observed on experimentally accessible time scales. As can be seen in Figs.~\ref{fig:fig4}b-c, in the thermal phase we observe significantly faster decay of the subsystem purity and faster growth in the Von-Neumann entropy as compared to the DTC phase.

Finally, we move to finite size scaling analysis to estimate the phase transition point ($g_c$) extracted from the generalized Edward-Anderson order parameter~\cite{Parameswaran_localization}  $\chi_{\text{EA}}= \frac{1}{N}\sum_{i \neq j} |\langle \hat Z_{i}^\dag \hat Z_{j}\rangle|^2$. This parameter measures glassiness from spin correlations across the chain and as a result, increases with $N$ in the MBL regime and decreases with $N$ in the thermal regime until it vanishes. Consequently, we can experimentally identify an approximate crossing that defines the phase transition point between the two phases. We observe this expected behavior by performing our Floquet circuits across a range of kicking strength $g$, and Floquet cycles. In Fig.~\ref{fig:fig4}c, we present the results in spin chains of length $N=6,9,12$ from averaging $\chi_{EA}$ across 5-10 Floquet cycles with 13 initial trit string instances. We observe an approximate critical kicking strength between the DTC and thermal phases in the range of $0.82 \lesssim g_c \lesssim 0.89$, consistent with the period tripling response observed in Figs.~\ref{fig:fig2},~\ref{fig:fig3} and our numerics (for more information see the Supplementary Materials).

\subsection*{Subspace Time Crystal}

Finally, we demonstrate that native qutrit operation also offers a platform to investigate time crystalline behavior within a subsystem of our model --- in our case a \textit{qubit} time crystal oscillating within the larger qutrit state space. In particular, we engineer an interpolation between a $\mathbb{Z}_2$ and $\mathbb{Z}_3$-DTC by adjusting our kicking gates. We select an initial trit string state with all qutrit population in the $\{ \ket{1},\ket{2}\}$ subspace and modify the global kick $\bar{X}^g  \rightarrow \bar X_{12}^{1-g} \bar X^g$, with $X_{12}$ being the qubit Pauli $X$ gate in the $\{ \ket{1},\ket{2}\}$ subspace, leading to effective disordered $\mathbb{Z}_2$ Ising behavior for low $g$ and full $\mathbb{Z}_3$ CCM for strong $g$. As can be observed in the corresponding $\expval{M}$ FFT (Fig.~\ref{fig:fig5}a), for a small $g$, the population in the $\{ \ket{1},\ket{2}\}$ subspace responds as a $\mathbb{Z}_2$ DTC with characteristic subharmonic period doubling even as $g$ tunes away from the trivial point $g=0$ (Fig.~\ref{fig:fig5}b), before melting into a thermal state, and finally recovering, with a large $g$, a full $\mathbb{Z}_3$-DTC (Fig.~\ref{fig:fig5}c). Similar to the role of chirality in our driven CCM, this subspace time crystal behavior has no analog in qubit Ising systems.

\section*{Discussion}

The $\mathbb{Z}_3$ DTC and related driven CCM physics demonstrated in this work establishes native qudit hardware as a qualitatively distinct platform for exploring non equilibrium phases of matter. The chirality-controlled melting of DTC order, with no $\mathbb{Z}_2$ analog, exemplifies this distinction: the chiral angle is an intrinsic feature of the 
CCM that fundamentally governs domain-wall dynamics, spectral degeneracies, and DTC phase stability. Moreover, our experimental platform allows for this angle to be continuously controlled in-situ, revealing direct experimental observation of its effect on time-crystalline order. In a similar vein, our ability to engineer a $\mathbb{Z}_2$ DTC within our higher spin model points to a wide range of unexplored dynamical orders leveraging fractional --- or subspace --- kicking schemes.

Our $\mathbb{Z}_3$ system is the simplest member of a family of $\mathbb{Z}_n$ chiral clock models with progressively richer phase diagrams, capable of hosting exotic features such as parafermionic edge modes with braiding statistics beyond those of Majorana zero modes~\cite{Fendley_2012, Fendley_stability,Moessner}. 
As superconducting qudit processors scale to larger system sizes and higher local dimensions and benefit from improved control schemes, these phases become within reach of near-term quantum simulators. Additionally, the native qutrit operations and tunable interactions demonstrated in this work provide a natural setting for realizing qutrit topological codes~\cite{iqbal2025qutrit} and exploring non-Abelian anyon physics in a solid-state setting. Like the $\mathbb{Z}_3$ DTC itself, these directions stand to benefit substantially from leveraging the full structure of the qudit Hilbert space.

\section*{Acknowledgments}
NG, NS thank Siddharth Parameswaran and Thomas Iadecola for fruitful discussions. This material was funded by the U.S. Department of Energy, Office of Science, Office of Advanced Scientific Computing Research Quantum Testbed Program under contract DE-AC02-05CH11231. NS and WAdJ acknowledge support from the “Embedding Quantum Computing into Manybody Frameworks for Strongly Correlated Molecular and Materials Systems” project, by the U.S. DOE, Office of Science, Office of Basic Energy Sciences, Division of Chemical Sciences, Geosciences, and Biosciences. SA acknowledges support from the Army Research Office (Grant No. W911NF-24-1-0079) and the Harvard Quantum Initiative.
NYY acknowledges support from the U.S. Department of Energy, Office of Science, National Quantum Information Sciences Research Centers, Quantum Systems Accelerator.
JEM acknowledges support from the U.S. Department of Energy, Office of Science, National Quantum Information Sciences Research Centers, Quantum Science Center.

\putbib[Z3Driven]

\end{bibunit}

\clearpage
\onecolumngrid

\begin{bibunit}
\begin{center}
{\large\bfseries Supplemental Material: A Qutrit Time Crystal in a Native Chiral Clock Model}
\end{center}

\vspace{1em}
\makeatletter
\let\addcontentsline\origaddcontentsline
\makeatother

\tableofcontents



\hypersetup{pdflang={English},colorlinks=true,linkcolor=Cerulean,citecolor=RoyalBlue,urlcolor=RoyalBlue,breaklinks=true}

\newcommand{\s}{\hat{Z}}
\renewcommand{\t}{\hat{X}}

\newpage

\section{Qutrit Device Specs and Operation}
\subsection{Coherences and Frequency Allocation}

\begin{figure}[h!]
    \centering
    \includegraphics[width=\linewidth]{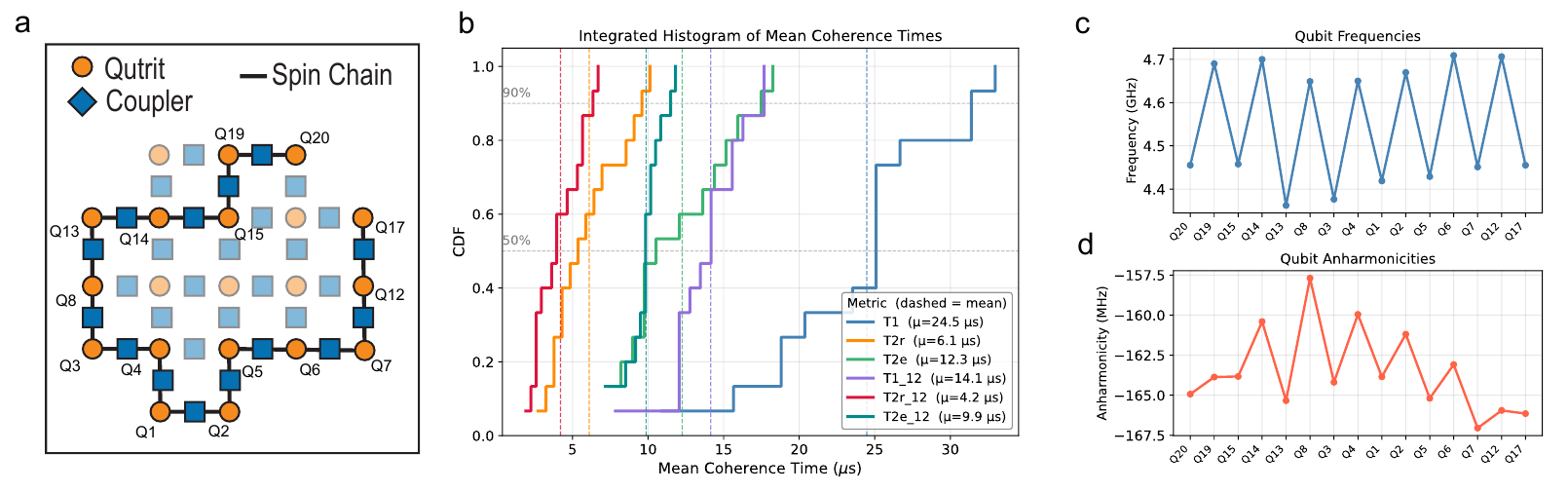}
    \caption{\textbf{Qutrit Device Specs} (a) Layout of 20 qubit quantum processor utilized in our DTC experiments. 1D spin-chains are selected from the square lattice topology. Actively pulsed qubits and couplers are shown in bold, while idling qubits and couplers are faded. (b) Qutrit coherence statistics across the 15 qutrit spin-chain where all means are collected from averaging over 100 separate coherence measurements. (c) Qubit ($\omega_{01}/2\pi$) frequencies across the spin-chain. (d) Qubit anharmonicities across the spin-chain. }
    \label{fig:coherences}
\end{figure}

The quantum processor employed in our DTC experiments consists of a 2D grid of 20 superconducting flux-tunable transmon qubits with interactions mediated by 30 flux tunable couplers designed, fabricated, and packaged by IQM \cite{abdurakhimov2024technologyperformancebenchmarksiqms}. The processor is housed on premise at UC Berkeley in a Blue-Fors XLD dilution refrigerator, employing custom wiring with special integrated components such as a cryogenic bias-tee array for the qubit flux lines supplied by YQuantum (YQ-CBT12-1G) and TWPAs provided by Silent Waves as well as HERD (High Energy Radiation Drain) filters on the readout lines. The RF control hardware used in our experiments is the open-source qubic system developed at Lawrence Berkeley National Laboratory \cite{qubic}.

As discussed in the main text, we perform qutrit based computation on our 20-qubit processor; for our DTC experiments, we select sub-chains within the lattice, such as the 15 qutrit chain presented in Fig.~\ref{fig:coherences}a. The transmons are flux-biased to the top of their tuning curves (unless there is a parasitic two-level system spectrally nearby) to maximize coherence during circuit execution. Statistics on the relevant T1, T2 Ramsey, and T2 echo coherence times between the $\ket{i},\ket{i+1}$ levels of our transmon qutrits are shown in Fig.~\ref{fig:coherences}b. Finally, in Fig.~\ref{fig:coherences}c,d we present the frequency and anharmonicity allocation along the 15 qutrit chain used in our largest DTC experiment from the main text. We next provide further information on single qutrit gate and readout operation on the device, with information on engineering the CCM interaction provided later in this supplement. 

\subsection{Single Qutrit Gates}

\begin{figure}[h!]
    \centering
    \includegraphics[width=\linewidth]{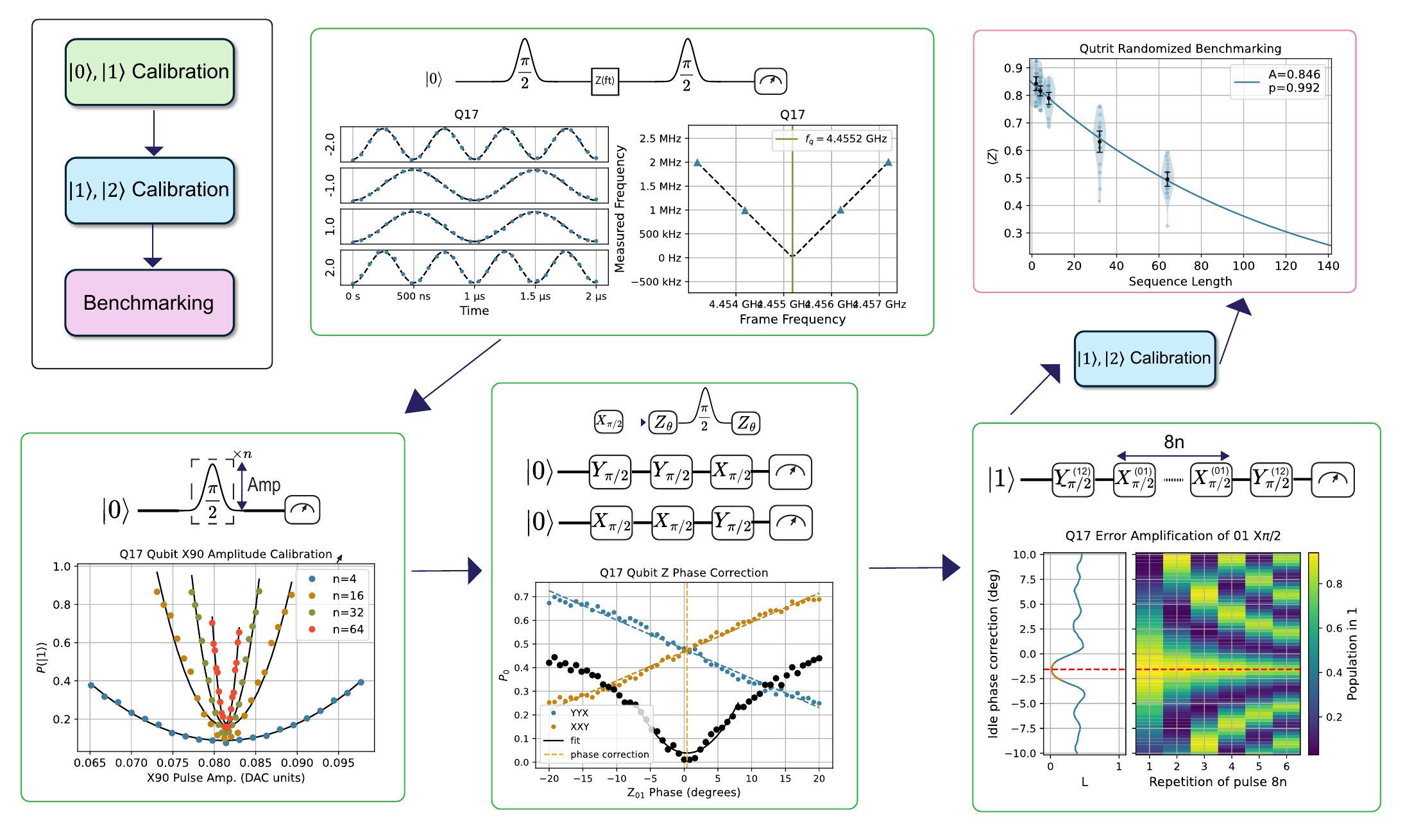}
    \caption{\textbf{Single Qutrit Gate Calibration} (black) Calibration schematic workflow. We calibrate arbitrary qutrit operations by first calibrating the 01 subspace of the qutrit, before calibrating the 12 subspace, and finally benchmarking the gates. (green) The pulse/circuit diagrams and calibration results for the subspace qubit calibrations. (blue) The same calibration routines modified to apply to the 12 subspace of the qutrit are then performed. (purple) The calibration can finally be verified by performing qutrit randomized benchmarking, where the Clifford gates are decomposed into 6 rounds of subspace $\pi/2$ rotations as well as virtual Z gates.}
    \label{fig:qutrit-gates}
\end{figure}

In our implementation of the driven CCM, our local ``kicking" is performed via single qutrit gates realized from microwave driven transitions between the $\ket{i}\leftrightarrow\ket
{i+1}$ levels of the transmon. In the experiment, we calibrate $X_{\pi/2}$ pulses in the $\ket{0},\ket{1}$ and $\ket{1},\ket{2}$ subspaces of each transmon qutrit, and employ virtual Z gates \cite{PhysRevA.96.022330} as our continuous degree of freedom. By working in the rotating frame of the qutrit, i.e., the frame tracking both the $\ket{0},\ket{1}$ and $\ket{1},\ket{2}$ transitions such that we accumulate no phase while idling, phase updates to the two driven transitions generate virtual Z gates of the form,
\begin{equation}
    Z(\theta_{01},\theta_{12}) = \text{diag}(1,e^{i\theta_{01}}, e^{i\theta_{12}}).
\end{equation}
We calibrate subspace qutrit $\pi/2$ rotations using well-known techniques from qubit calibration (see Fig.~\ref{fig:qutrit-gates}), essentially treating each qutrit subspace as an individual qubit for calibration purposes. We bootstrap the calibration beginning with calibration of the 01 subspace of the qutrit. We find the relevant transition frequency via a frequency-swept Ramsey experiment. Then the pulse amplitude is calibrated using standard error amplification techniques. Finally, as in Ref.~\cite{PhysRevLett.126.210504}, we calibrate a phase correction on the idling qutrit subspace from the Stark shift due to the pulses in the driven qutrit subspace. Once the calibration is completed in one subspace, the same suite of calibrations is repeated in the other subspace.

Given a calibrated $X_{\pi/2}^{(01)}$ and  $X_{\pi/2}^{(12)}$ gate and a continuous degree of freedom via our virtual Z gates in both qutrit subspaces, we can generate arbitrary single qutrit gates using a decomposition into 3 rounds of subspace SU(2) rotations (also known as Givens rotations) \cite{PhysRevA.97.022328}. We subsequently benchmark the calibration of our qutrit subspace rotations using randomized benchmarking over the full qutrit Clifford group \cite{PhysRevLett.126.210504}.

\begin{figure}[h!]
    \centering
    \includegraphics[width=\linewidth]{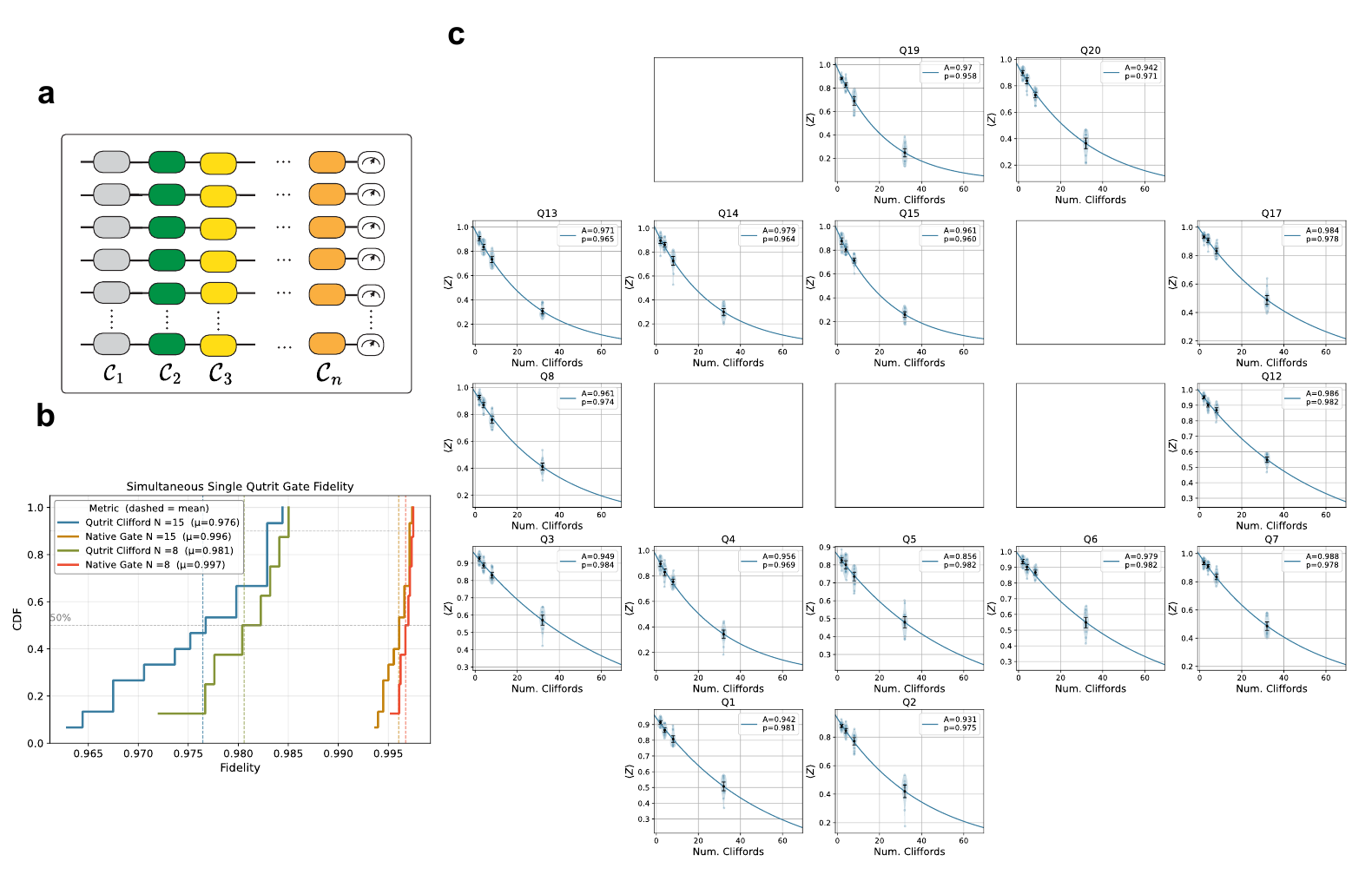}
    \caption{\textbf{Simultaneous Randomized Benchmarking on a Qutrit Chain.} (a) Circuit diagram of simultaneous randomized benchmarking circuits. Each $\mathcal{C}_i$ consists of a string of randomly chosen qutrit Clifford gates for each qutrit in the chain. $n-1$ Cliffords are performed before a final cycle of qutrit gates $\mathcal{C}_n$ which inverts all the previous Clifford gates. (b) Integrated histogram of simultaneous qutrit randomized benchmarking results on an $N=15$ and $N=8$ chain. We compare both the simultaneous single qutrit gate fidelity, as well as the average fidelity of the native gates (recall each Clifford is composed of 6 native pulses). (c) Qutrit randomized benchmarking results on the $N=15$ chain employed in our largest DTC experiment in the main text.}
    \label{fig:supp-rb}
\end{figure}

In our experiment, we parallelized the calibration routines in Fig.~\ref{fig:qutrit-gates}, and perform all the calibrations simultaneously across the chain of qutrits we plan to use in our driven CCM. We show the results of benchmarking a 15 qutrit chain in Fig.~\ref{fig:supp-rb}c. We observe that crosstalk degrades the overall performance of our qutrit operations as gate fidelities decrease as we increase the size $N$ of our qutrit chain (see Fig.~\ref{fig:supp-rb}b). We attribute this partially to residual $ZZ$ coupling which for the strongest cross-Kerr term between our qutrits remains around 25 KHz on average (for more information see the tunable coupler section of the supplement) but mainly to microwave crosstalk. Future work to significantly improve simultaneous qutrit gate operation could employ crosstalk compensation pulses or explore advanced frequency allocation and pulse shaping techniques such as were developed in Ref.~\cite{wesdorp2026mitigatingcrosstalkerrorssimultaneous}.

\newpage

\subsection{Qutrit Readout Characterization}
\begin{figure}[h!]
    \centering
    \includegraphics[width=\linewidth]{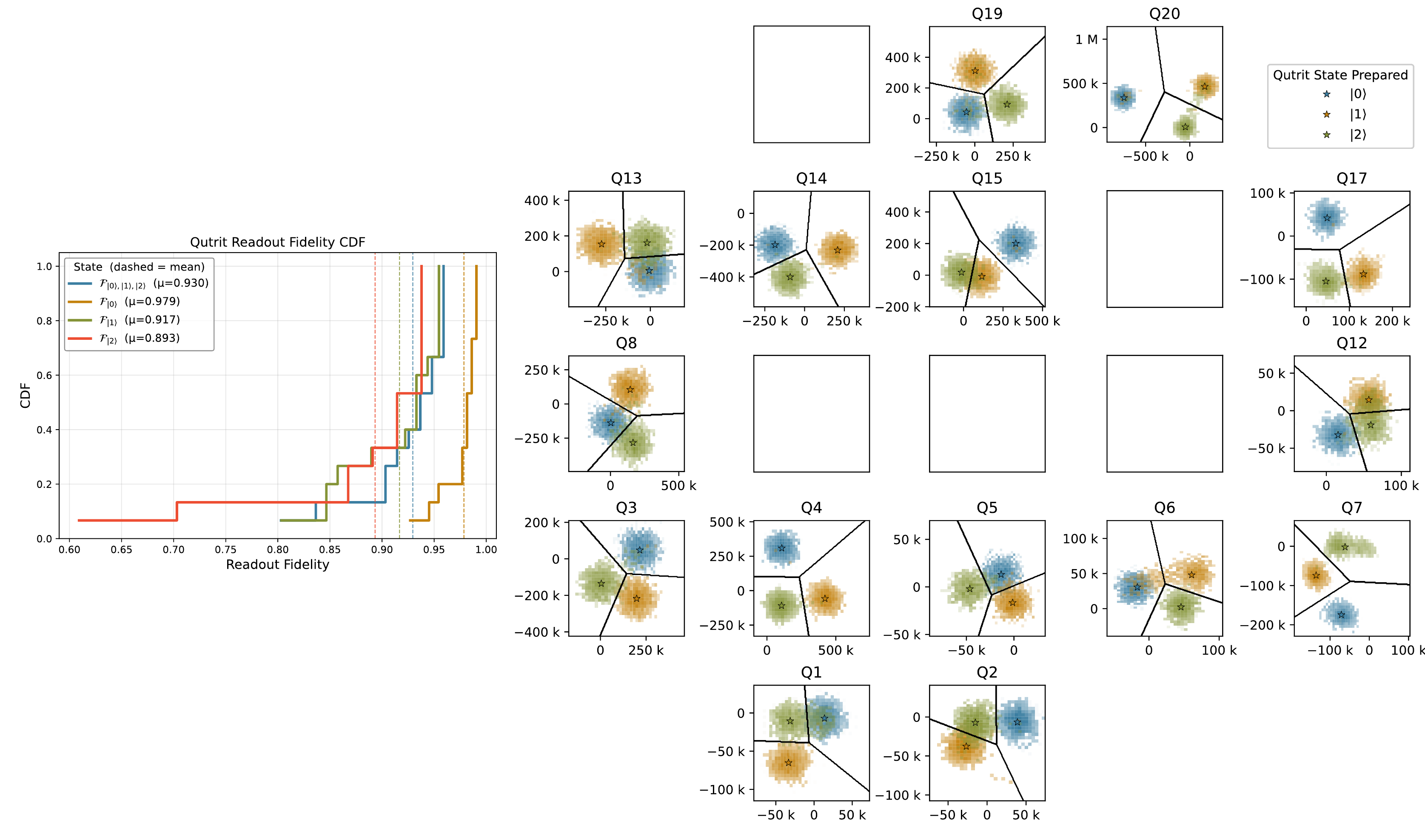}
    \caption{\textbf{Qutrit Readout Characterization} (left) Statistics of qutrit readout fidelities across the 15 qutrit chain used in the largest DTC experiment in the main text. (right) The histogram blobs in the IQ plane of our dispersive qutrit readout with 2048 shots per prepared qutrit state $\ket{i}$, $i \in \mathbb{Z}_3$. Results with $\ket{0}$,$\ket{1}$, and $\ket{2}$ prepared are shown respectively in blue, orange, and green.}
    \label{fig:supp-readout}
\end{figure}

Finally, we characterize the 3 level readout of our transmon device. We perform readout dispersively, applying a microwave pulse to probe our readout resonators for 500ns, with the amplitude and frequency of the readout pulse chosen to distinguish all 3 qutrit states of the transmon simultaneously. In Fig.~\ref{fig:supp-readout} we present the readout fidelities and integrated dispersive readout signals for qutrit readout across the full 15 qutrit chain employed in our largest DTC experiment. Notably, as our experiments only required a terminal readout and no mid-circuit measurement, we did not maximize the fidelity of our readout operations, and considerable room for improvement exists with these fidelities. Additionally, we do not here characterize the QNDness of the readout, which would be critical to understand in a setting such as a qutrit quantum error correction code demonstration. Nonetheless, we report an average qutrit readout fidelity of 93\% across our device.

\newpage
\subsection{Tunable Coupler Operation}
\subsubsection{Qutrit Spin-Spin Interactions}
As discussed in the main text, by encoding qutrits in our transmon lattice, we are able to natively realize a driven CCM. In this section, we provide more information on our tunable coupler method for realizing the interaction terms in the CCM performed in the main text. Specifically, we describe how these interactions are generated from tunable coupler pulses and show the freedom to tune the strength, $J$, and chiral angle, $\theta$, in the CCM interaction.

\begin{figure}[h!]
    \centering
    \includegraphics[width=\linewidth]{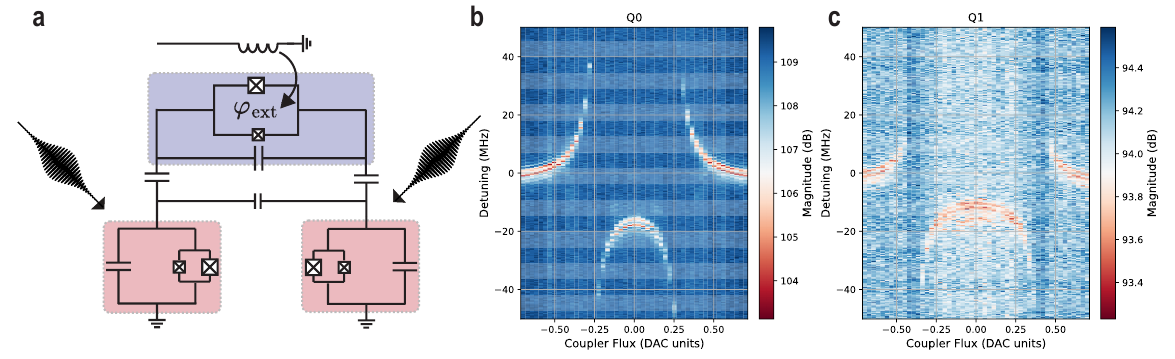}
    \caption{\textbf{Qubit two-tone spectroscopy vs. coupler flux} (a) Circuit diagram with two transmon qubits (red) coupled to a third flux tunable transmon coupling qubit (blue). The qubits are driven with microwave pulses and the coupler's frequency is adjusted with flux. (b-c) Q0/Q1 qubit two-tone spectroscopy vs coupler flux. Note that, as can be observed from the avoided crossing, the coupler tunes from above the qubits at the integer flux to below the qubits at its half flux point.}
    \label{fig:supp-twotone}
\end{figure}

We realize a driven effective Hamiltonian that locally maps to the spin-spin interaction term in the CCM by applying fast flux pulses on the tunable couplers which mediate hybridization between neighboring transmons on our chip. In practice, this works by choosing an idling flux point where the tunable coupler mediated hybridization is small, and then pulsing the coupler to a regime where the dispersive coupling between qutrits is large by decreasing the detuning between the coupler and the qutrit transition energies. As shown via the qubit two-tone vs coupler flux experiment in Fig.~\ref{fig:supp-twotone}, the coupler at 0 flux idles above the two qubit frequencies, before being tuned through them (see the avoided crossing).

Our unit-cell (for more information see Ref.~\cite{PRXQuantum.4.010314}) consists of two transmons modeled as duffing oscillators, coupled via a third transmon acting as a tunable coupler. The data transmons have some direct capacitive coupling ($g_{12}$) and are coupled to the tunable coupler ($g_{ic}$). This realizes the system Hamiltonian,
\begin{equation}\label{eq:ck}
    H_{\text{sys}} = \sum_{i = 0,1} \omega_i a_i^\dag a_i + \frac{\eta_i}{2}a_i^\dag a_i ^\dag a_i a_i + \omega_c(\Phi_{\text{ext}})a_c^\dag a_c + g_{12}(a_1^\dag a_2 + a_1 a_2^\dag) + \sum_{i=1,2}g_{i,c}(a_i^\dag a_c + a_i a_c^\dag),
\end{equation}

where $a_i$, $\omega_i$, and $\eta_i$ are respectively the qubit bosonic annihilation operator, frequency, and anharmonicity of transmon $i$ and we have neglected the nonlinearity of the coupler for simplicity. As in Refs.~\cite{PhysRevLett.130.030603,PhysRevLett.125.240502,chen2026unlockingfastadiabaticcz}, we engineer a tunable $ZZ$ interaction between our transmons tuned by the coupler flux $\varphi_{\text{ext}}$, that when truncated to the coupled qutrit Hilbert space, realizes (up to local phases), a cross-Kerr Hamiltonian \cite{scrambling, goss2022high},
\begin{equation}
    H_{\text{CK}} = \sum_{i,j \in \{1,2\}}\alpha_{i,j}(\varphi_{\text{ext}})\ket{i0j}\bra{i0j}.
\end{equation}

\begin{figure}[h!]
    \centering
    \includegraphics[width=\linewidth]{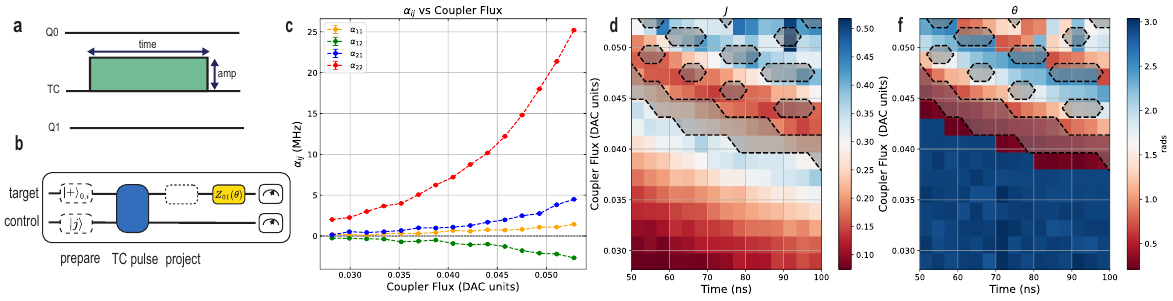}
    \caption{\textbf{Characterizing the qutrit spin-spin interactions} (a) Pulse diagram for implementing the spin-spin gates. The two qubits (Q0,Q1) are idle and a square baseband pulse is applied to the tunable coupler (TC). (b) A cicuit diagram of the qutrit generalized JAZZ experiment used to measure the cross-Kerr rates and entangling phases. (c) The extracted cross-Kerr rates as a function of coupler flux. (d-f) The extraced spin-spin gate parameters expressed in terms of the relevant CCM coeffecients $J$ and $\theta$ for our $\mathbb{Z}_3$-DTC. For example, we highlight regions in flux pulse amplitude and time where the spin-spin gates realize chiral angles that are greater than 0.3 rads detuned from degeneracy inducing points (i.e. $n \pi/3$).}
    \label{fig:supp-crosskerr}
\end{figure}

In the dispersive limit where the qubit-coupler detunings are larger than their coupling $g_{ic}$ the coupler remains unpopulated and we can use perturbation theory to compute the cross-Kerr ($ZZ$-like) terms as $\alpha_{i,j} = (E_{i,0,j} - E_{i,0,0}) - (E_{0,0,j} - E_{0,0,0})$, employing the energy labeling \{transmon 1, coupler, transmon 2\}. In practice, analytically capturing the behavior of the full qutrit dispersive coupling requires careful attention to higher-order perturbative corrections and is beyond the scope of this text \cite{fors2024comprehensiveexplanationzzcoupling}.

We measure these cross-Kerr term as a function of flux applied to the coupler in Fig.~\ref{fig:supp-crosskerr} using a generalized JAZZ sequence \cite{PhysRevLett.119.180501} (see Fig.~\ref{fig:supp-crosskerr}b), and display the $\alpha_{ij}$ rates in Fig.~\ref{fig:supp-crosskerr}c. Experimentally we find the statically decoupled point where $ZZ$ is minimized corresponds to when the coupler is significantly below the qubits in frequency. Notably, as the qubits are negatively anharmonic and the coupler frequency is below the qubits, the $\alpha_{22}$ rate increases dramatically faster as a function of coupler flux than the other cross-Kerr terms. 

We can subsequently extract the qutrit conditional phases as a function of flux and time from the JAZZ experiments. Upon doing so, we map the spin-spin gates $U = e^{-iH_{\text{ck}}t}$ to the two-body interaction term in the CCM where the angles $\theta_{ij}= 2\pi \times \alpha_{ij}t$ are used to calculate $\{J,J',\theta,\theta' \}$ (see the supplementary section concerned with mapping the cross-Kerr to the CCM Hamiltonian). Via this mapping, as discussed in the main text, we can generate spin-spin gates that either lead to stable or unstable DTC behavior via the tuning of the chiral phase. In Fig~\ref{fig:supp-crosskerr}d,f we show regimes in the space of coupler flux pulse amplitude and duration which generate two-body spin-spin gates that lead to stable DTC behavior. Such characterization can be performed across all coupling sites in the spin chain to generate either stable or unstable DTC behavior as demonstrated in the main text.

\subsubsection{Calibration of Tunable Coupler for Isolated Single Qutrit Operation}

\begin{figure}[h!]
    \centering
    \includegraphics[width=\linewidth]{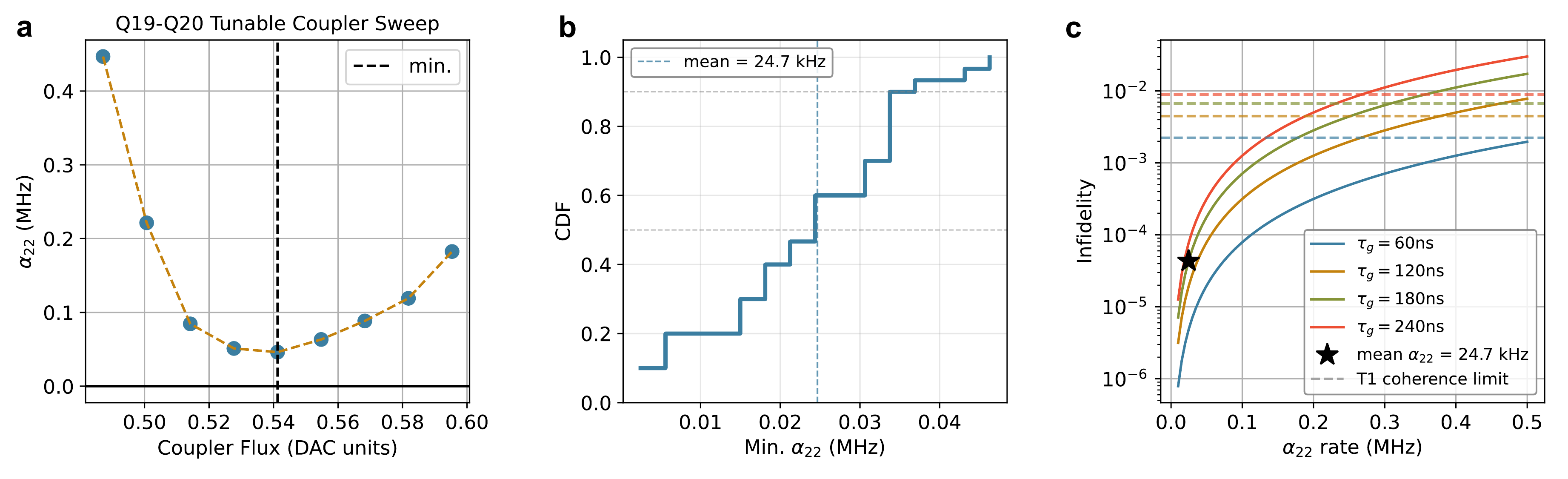}
    \caption{\textbf{Calibrating the Tunable Coupler Idling Flux Point} (a) An example $\alpha_{22}$ vs coupler flux biasing experiment, where the coupler flux is swept to find a minimal cross-Kerr coupling rate. (b) The extracted minimal $\alpha_{22}$ cross-Kerr rate across all 30 qutrit-qutrit pairs on our 20 transmon processor. We find a mean nulling of 24.7 KHz averaged over all 30 tunable couplers. (c) Error budgeting simultaneous single qutrit gates between two qutrits as a function of gate time ($\tau_g$) for cross-Kerr unitary errors (solid lines) and decoherence (horizontal dashed lines). The coherence limits are calcluated using the average T1 times across all qutrits from our statistics displayed in Fig.~\ref{fig:coherences}.}
    \label{fig:supp-couplerbias}
\end{figure}

In the last section, we demonstrated that the $\alpha_{22}$ cross-Kerr term grows much faster as a function of flux than the other $\alpha_{ij}$ terms owing to the negatively anharmonic nature of our transmon qutrits and the fact the coupler nulled $ZZ$ point is below the qutrit transitions in frequency. With this in mind, our strategy for finding the best idling coupler flux point for our qutrits is to null the $\alpha_{22}$ cross-Kerr term using the same JAZZ experiment from Fig.~\ref{fig:supp-crosskerr}. In Fig.~\ref{fig:supp-couplerbias}a, we show the results for a tunable coupler flux calibration experiment, where we sweep the coupler flux to find a point with minimal $\alpha_{22}$ cross-Kerr. As the majority of our qutrits are not straddling, we do not generally find an $\alpha_{22} = 0$ cancellation point, but notably, as demonstrated in Fig.~\ref{fig:supp-couplerbias}b, we can achieve an average $\alpha_{22}$ nulling of 24.7 KHz averaged across all 30 qutrit-qutrit coupled pairs on our 20 data qutrit processor. Finally we suggest a simple error budget for our simultaneous single qutrit gate operation; we estimate the error on our single qutrit operations by defining
\begin{equation}
    \mathcal{F} = \text{Tr}[U_{\text{id}}^\dag U_{\text{err}}]/d^2,
\end{equation}
where we define $U_{\text{err}} = e^{-i(2\pi \alpha_{22}\times t \ket{22}\bra{22} )}$, only keeping the $\alpha_{22}$ term as it dominates and neglecting the other cross-Kerr terms for simplicity. We can then calculate an estimate for the single qutrit gate error as a function of both gate time $\tau_g$ and cross-Kerr strength $\alpha_{22}$.

In Fig.~\ref{fig:supp-couplerbias}c, we display the cross-Kerr limited gate fidelities as a function of $\alpha_{22}$ rate. Notably, given our cross-Kerr nulling, for our arbitrary SU(3) gate time of 180 ns, gate fidelities over 99.9\% should in principle be achievable. We also plot the associated T1 limited gate fidelities given $\tau_g$, using the decoherence gate infidelity model from Ref.\cite{PhysRevLett.126.210504}. These results strongly suggest that by integrating tunable couplers, simultaneous single qutrit gate operations are no longer limited by quantum crosstalk, unlike in prior fixed coupling experiments \cite{goss2022high, goss_extending_2024, PhysRevLett.134.050601}. This now highlights that gate time remains a major bottleneck; in the near-term, future works can explore speeding up the single qutrit gate times by exploring multi-chromatic and optimal control techniques \cite{champion_multifrequency_2025,seifert_timeefficient_2022}.

\section{Extended Qutrit DTC Data and Characterization}

\subsection{Interactions Off}
\begin{figure}[h!]
    \centering
    \includegraphics[width=\linewidth]{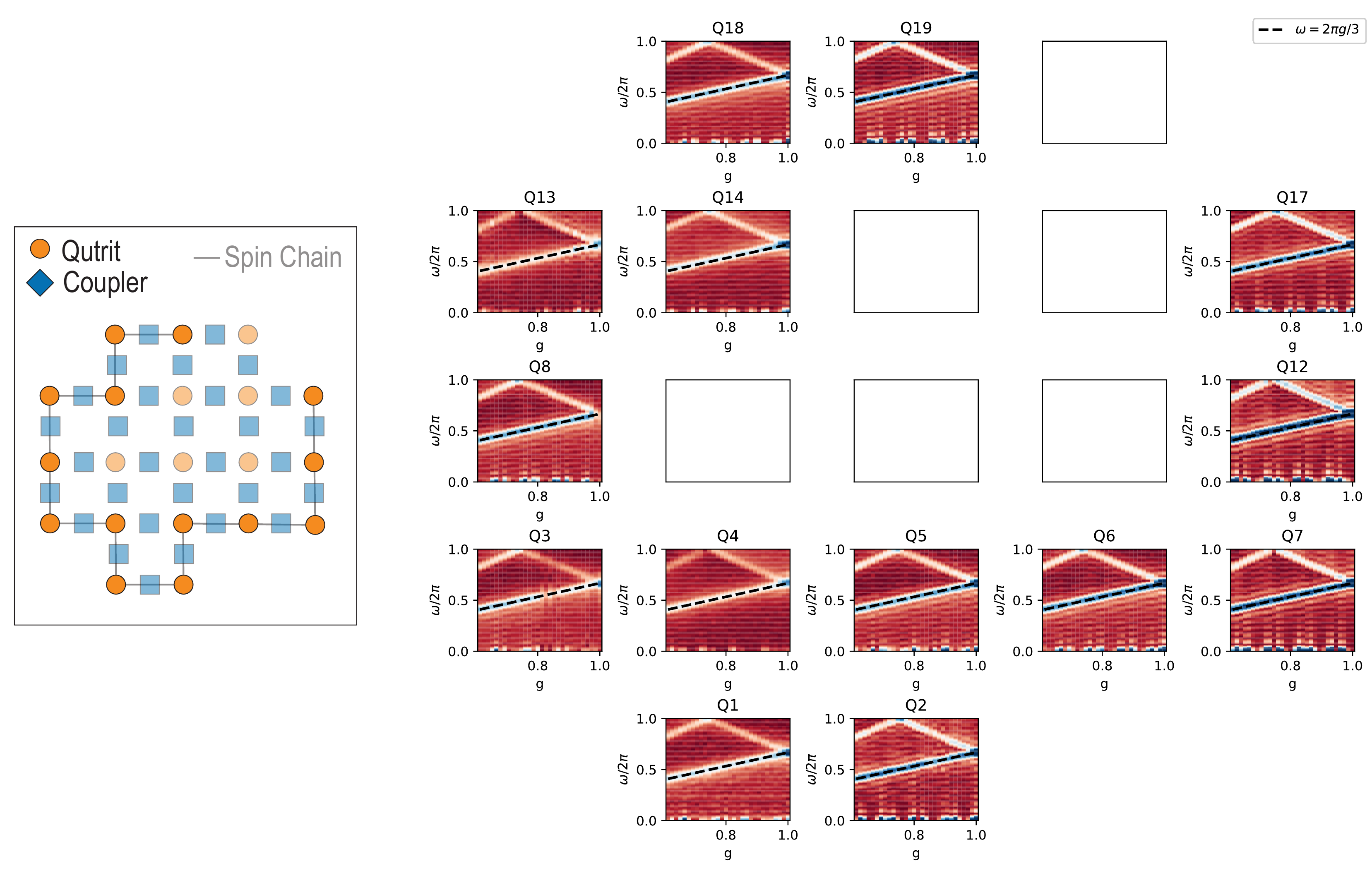}
    \caption{\textbf{Magnetization Spectroscopy with Interactions Turned Off.} (left) The chosen 14 qutrit spin chain on our 20 qutrit processor where the couplers are opaque to indicate the interaction Hamiltonian has been turned off. (right) The $\expval{M}$ FFT response across each site of the chain now displays trivial single particle responses to the kick $\bar{X}^g$ with all sites responding at $\omega = 2\pi g/3$ (black dashed line).  }
    \label{fig:supp-interaction-off}
\end{figure}
In Fig.2 in the main body we showed how the FFT of the averaged spin-1 magnetization $\expval{M}$ could serve as a useful probe of the transition between the thermal and $\mathbb{Z}_3$ DTC regimes of our Floquet model. Here we provide an effective ``null" experiment for this FFT by analyzing a chain of 14 qutrits with the couplings turned off. Notably, as depicted in Fig.~\ref{fig:supp-interaction-off}, as the kicking strength $g$ in $\bar{X}^g$ is tuned, the $\expval{M}$ FFT response across all sites follows the expected trivial single particle behavior at $\omega = 2\pi g/3$. As should be expected with no spin-spin interactions, the system here does not show any signatures of fast thermalization or robust period tripling. This $\expval{M}$ FFT response across the chain without interactions can now serve as a useful contrast to the experiments provided in the figure in the main text and the following section.

\newpage

\subsection{Full 15 Qutrit Chain Spectroscopy}

\begin{figure}[h!]
    \centering
    \includegraphics[width=\linewidth]{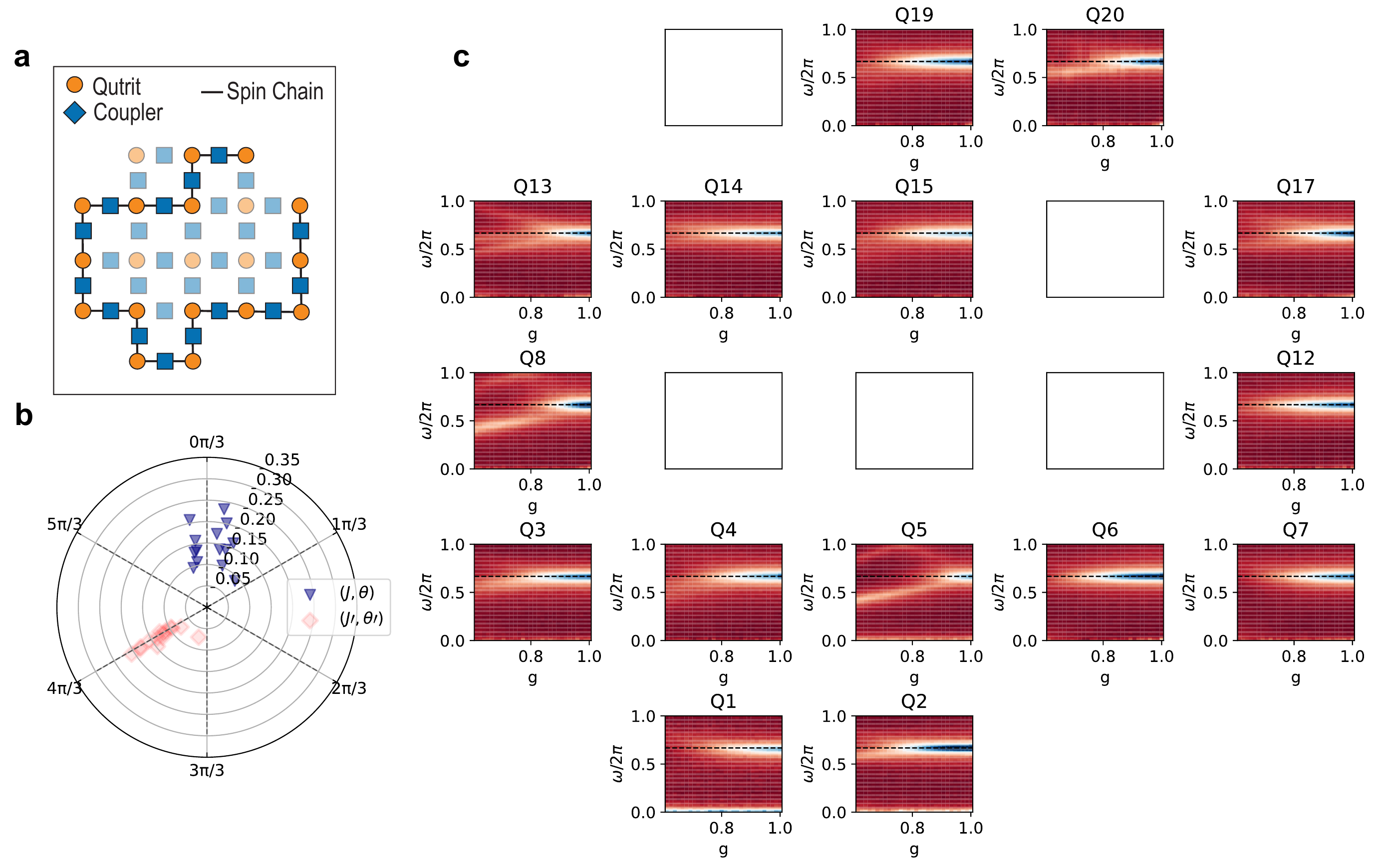}
    \caption{\textbf{Full 15 Qutrit Spin Chain Spectroscopy}. Extended results from the 15 qutrit $\mathbb{Z}_3$ DTC experiment performed in Fig. 2 in the main text. \textbf{a}, The 15 qutrit spin chain embedded in the lattice of our processor. \textbf{b}, The couplings $(J_j, J'_j, \theta_j, \theta'_j)$ across the qutrit chain. \textbf{c}, The $\expval{M}$ FFT response across the whole 15 qutrit chain. }
    \label{fig:fullchainDTC}
\end{figure}

We now provide extended data from the 15 qutrit $\mathbb{Z}_3$ DTC experiment depicted in Fig. 2 in the main text. In that experiment, we showed in the main text a transition from thermalization at $g < g_c$ to robust period tripling at $g>g_c$ by displaying the marginalized spin-1 magnetization $\expval{M}$ FFT response at a single site in our chain. Here we show the specific qutrit chain within our lattice (Fig.~\ref{fig:fullchainDTC}a), couplings across the 14 spin-spin interactions (Fig.~\ref{fig:fullchainDTC}b), and average spin-1 magnetization $\expval{M}$ FFT across the entire chain (Fig.~\ref{fig:fullchainDTC}c). Notably, we can observe that for a weak enough kicking strength $g$, every single FFT shows the characteristic response of rapid thermalization, whereas for a strong enough $g$ we recover robust period tripling. This $\mathbb{Z}_3$ DTC period tripling behavior can be observed to be robust when tuning the kicking strength away from the trivial point at $g=1$ across all sites, confirming the many body nature of the realized $\mathbb{Z}_3$ DTC state presented in the main text.

\newpage

\subsection{Dynamical Decoupling}

\begin{figure}[h!]
    \centering
    \includegraphics[width=0.93\linewidth]{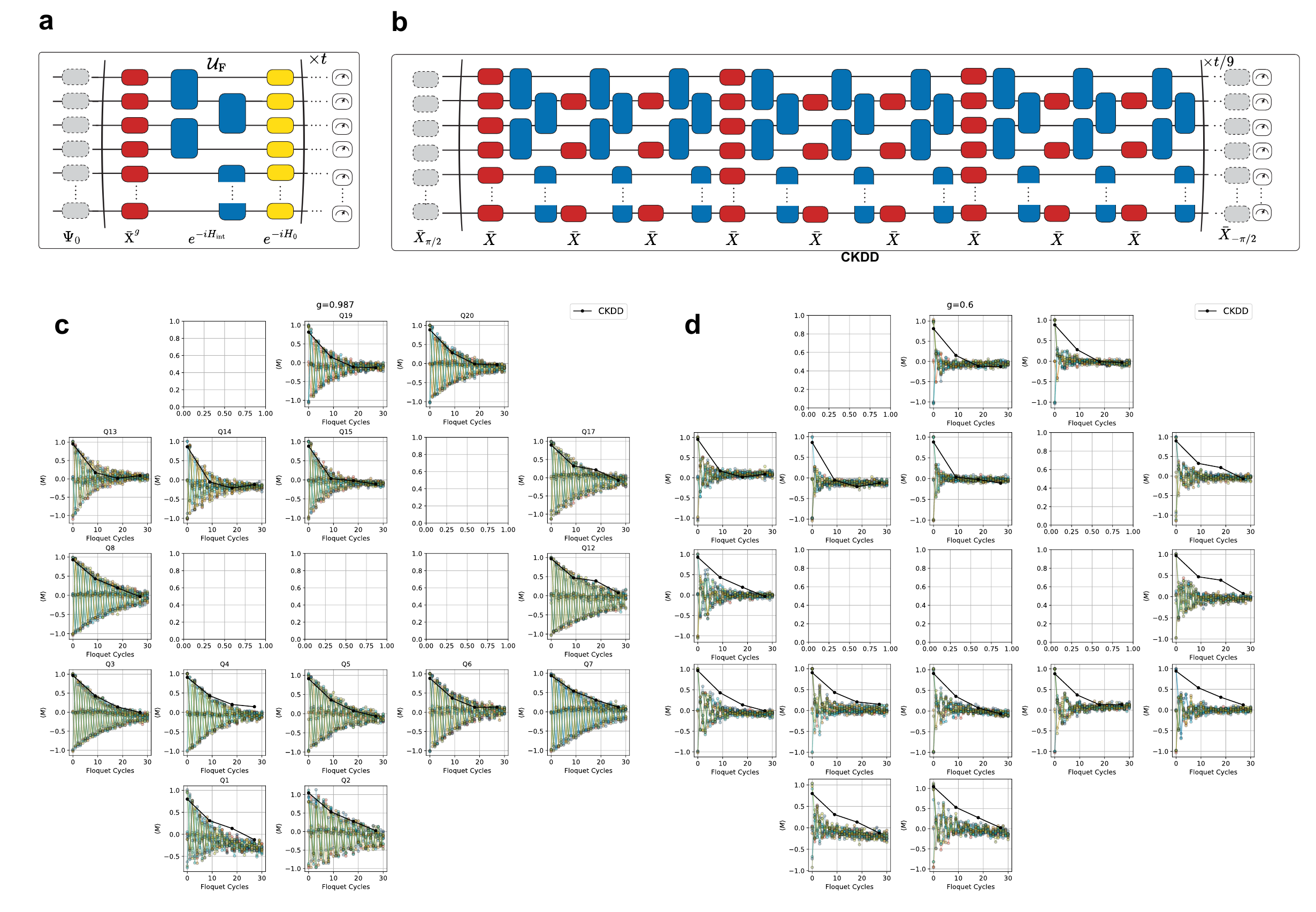}
    \caption{\textbf{Using Dynamical Decoupling to Demonstrate the DTC Oscillations are Primarily Limited by Decoherence.} \textbf{a}, The Floquet circuits employed to study the driven CCM physics. \textbf{b}, The cross-Kerr dynamical decoupling circuits we use to probe the role of decoherence in our DTC oscillations. $\ket{+} = \frac{1}{\sqrt{3}}(\ket{0} + \ket{1} + \ket{2})$ is prepared at each qutrit in the chain. We perform circuits with the same structure as the Floquet circuits, but now the kicking gates are replaced by full $\bar{X}$ rotations at each site. After 9 cycles of single and two-qutrit gates, the cross-Kerr evolution is removed and we can project the prepared states back onto $\ket{0}$ to assess the survival probability. \textbf{c}, Comparing the DTC oscillations for a strong kick of $g= 0.987$ in a 15 qutrit chain with the CKDD experiment survival probability. \textbf{d}, The same comparison in the thermal regime with $g= 0.6$. }
    \label{fig:supp-ckdd}
\end{figure}

We expect that in the $\mathbb{Z}_3$ DTC regime (i.e. for a strong kick $g> g_c$) the subharmonic period tripling response of our qutrit chain should be robust, with oscillations that are principally limited by system decoherence. In prior works studying qubit time crystals, such as Ref.~\cite{mi2022time}, the authors confirmed the stability of their $\mathbb{Z}_2$ time crystal by directly reversing the time evolution of their circuits and comparing the resulting envelope to their DTC oscillation envelope. To do this, the authors of that work engineered gates that exactly reversed the $ZZ$ or Ising couplings in their model. In principle, such a method could also be employed in our experiment, but it poses a very serious experimental challenge due to our spin-spin interactions now tuning 4 entangling phases (in contrast to the single phase in the qubit experiments). 

Instead, to estimate the role of decoherence in our circuits and compare it to our DTC experiments, we employ the cross-Kerr dynamical decoupling routine first introduced in Ref.~\cite{PhysRevLett.134.050601} (see Fig.~\ref{fig:supp-ckdd}b). To probe average case noise at each qutrit site, we prepare $\ket{+} = \frac{1}{\sqrt{3}}(\ket{0} + \ket{1} + \ket{2})$ at each site in the chain. We then perform circuits with the same structure as the Floquet circuits, but with the kicking gates replaced by full $\bar{X}$ rotations at each site. After 9 cycles of single and two-qutrit gates, the cross-Kerr(or spin-spin) time evolution is removed and we can project the prepared states back onto $\ket{0}$ to assess the survival probability. In general, consistent with our expectations, we find that the DTC oscillation decay rates closely match the CKDD experiments (Fig.~\ref{fig:supp-ckdd}c) whereas in the thermal regime the spin-1 magnetization decay is considerably faster than the bare system decoherence. 
\newpage

\subsection{Subspace Time Crystal}

\begin{figure}[h!]
    \centering
    \includegraphics[width=\linewidth]{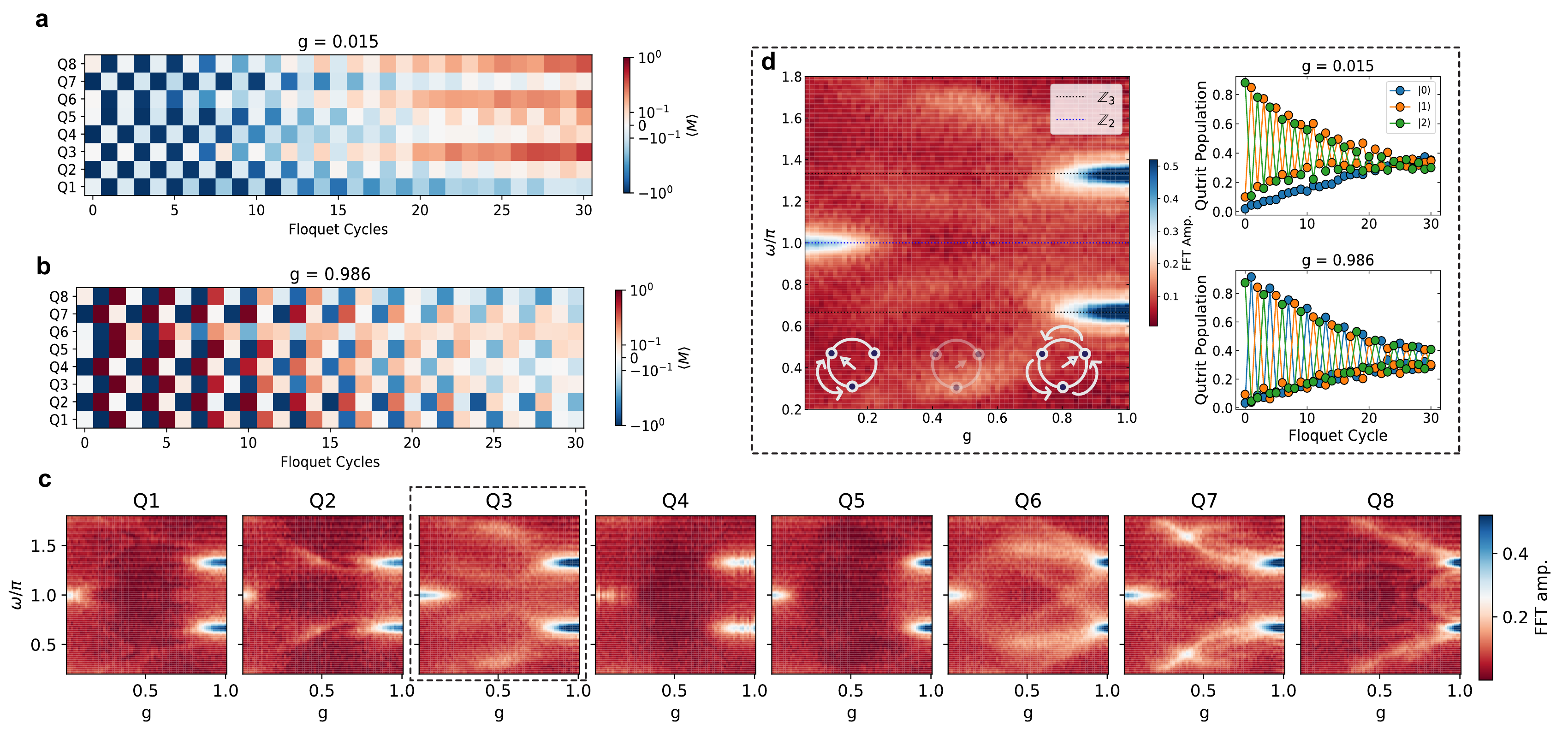}
    \caption{\textbf{Extended $\mathbb{Z}_2$ to $\mathbb{Z}_3$ DTC Data from Fig. 5 in the main text.} \textbf{a}, $\expval{M}$ response for the modified Floquet circuits with $g=0.015$. \textbf{b}, $\expval{M}$ response for the modified Floquet circuits with $g=0.986$. \textbf{c}, The FFT of the $\expval{M}$ response as a function of $g$ across the entire 8 qutrit chain employed in the experiment. \textbf{d}, Fig 5 from the main text, now identified with its position as one of the bulk sites within the 8 qutrit chain.}
    \label{fig:supp-z2toz3}
\end{figure}

As discussed in the main text, in addition to the chiral nature of our interaction Hamiltonian and its role in determining DTC stability, another distinguishing phenomena available in our higher spin model is the ability to access subspace DTC behavior. In particular, we modified our local kicking gates in our Floquet circuits from $\bar{X}^g \rightarrow \bar{X}^g\bar{X}_{12}^{1-g}$, where $X_{12}$ represents the qubit Pauli $X$ operator embedded in the $ \{ \ket{1}, \ket{2} \}$ subspace of the qutrit. We study these modified Floquet circuits on an 8 qutrit chain. As our subspace kicking gates now only rotate the $\{ \ket{1}, \ket{2}\}$ subspace, we prepare the initial trit string state $\ket{12121121}$ such that for small kicking strength $g$, the qutrit population at each site  in the spin chain should ideally oscillate within this subspace. 

As shown in Fig.~\ref{fig:supp-z2toz3}a, for a weak but non-trivial kick (i.e. $g\neq0$), we observe the characteristic subharmonic period doubling oscillations of a $\mathbb{Z}_2$ DTC in the $\{\ket{1},\ket{2} \}$ subspace across the entire chain. Similarly, as shown in Fig.~\ref{fig:supp-z2toz3}b, for strong kicking strengths $g$, we recover robust period tripling behavior consistent with a $\mathbb{Z}_3$ DTC. By observing the $\expval{M}$ FFT spectroscopy in Fig.~\ref{fig:supp-z2toz3}c, we can observe a line at $\omega=\pi$ visible at each site for weak $g$ corresponding to subharmonic period doubling, before observing the expected thermal melting in the intermediate kicking regime, and finally recovering the period tripling response at $\omega = 2\pi/3$ for large $g$. Notably, as can be seen in this spectroscopy, the period doubling, thermal melting, and period tripling responses were observable across all spin sites, signaling the many-body nature of the observed behavior.

\newpage
\subsection{Spin 1 Magnetization Spectroscopy Chiral Angle Dependency}
\begin{figure}[h!]
    \centering
    \includegraphics[width=\linewidth]{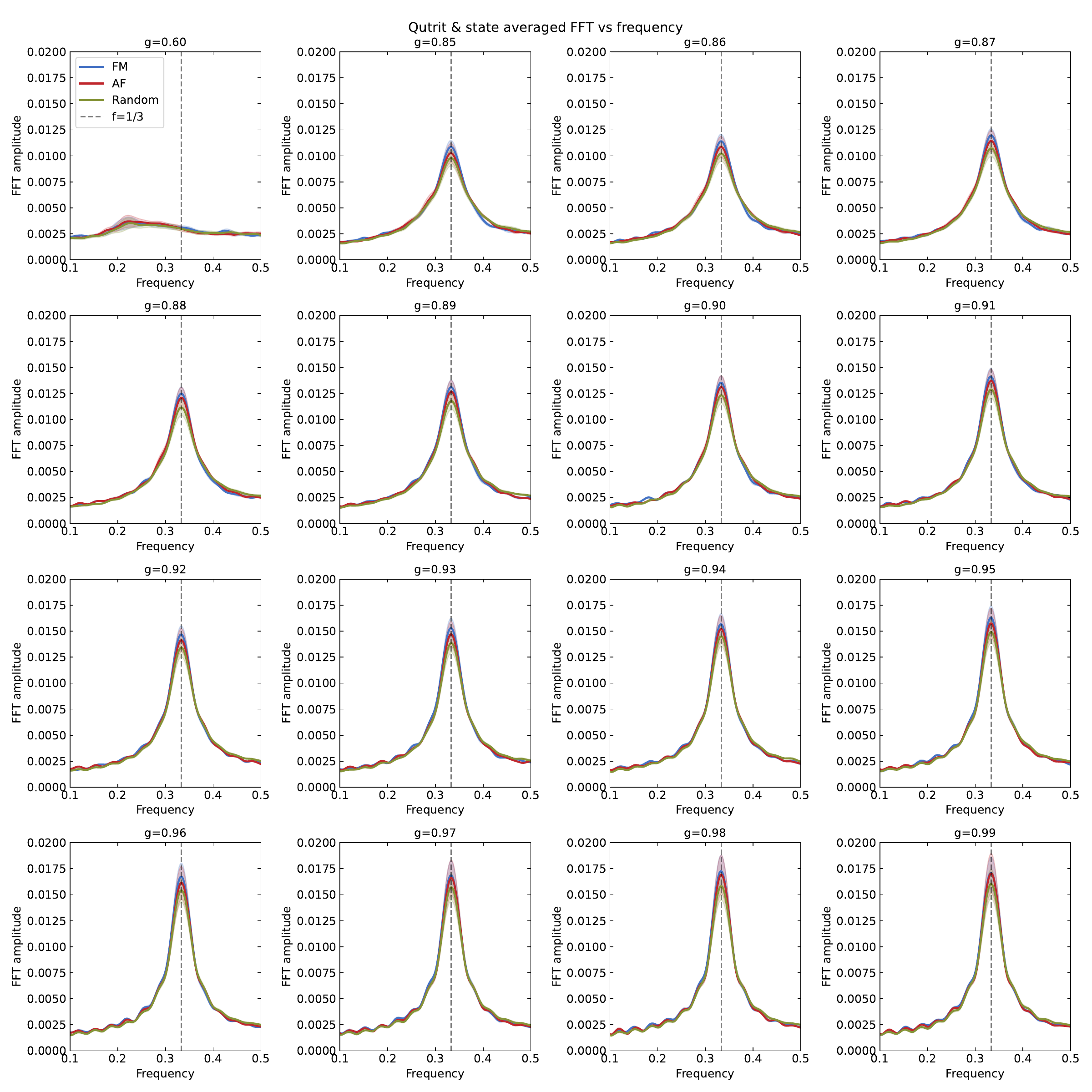}
    \caption{\textbf{Spin 1 Magnetization Spectroscopy vs $g$ for $\theta \neq 0$}. We present extended results from the experiment reported in Fig.~3 in the main text. Notably, with chiral couplings present in the system, the $\expval{M}$ FFT response shows period tripling subharmonic response across all initial states even for $g$ as low as 0.85.}
    \label{fig:fft-thetanonzero}
\end{figure}
\begin{figure}[h!]
    \centering
    \includegraphics[width=\linewidth]{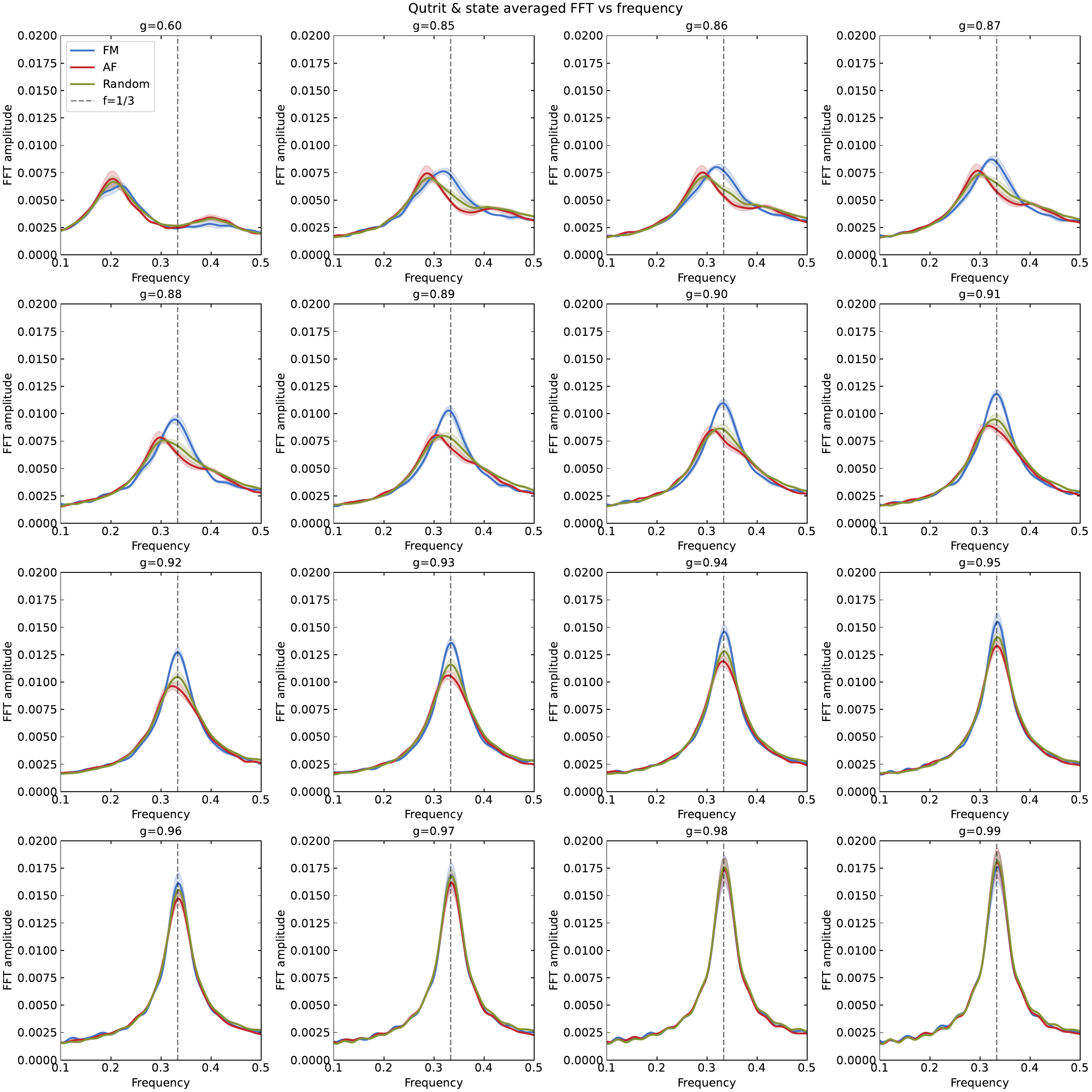}
    \caption{\textbf{Spin 1 Magnetization Spectroscopy vs $g$ for $\theta = 0$}. We present extended results from the experiment reported in Fig.~3 in the main text. Now with chirality turned off (or $\theta_j =0$), the $\expval{M}$ FFT response shows locked period tripling behavior for low $g$ only for the ferromagnetic states.}
    \label{fig:fft-thetazero}
\end{figure}
\newpage
\newpage

\subsection{Finite System Size Analysis}
\begin{figure}[h!]
    \centering
    \includegraphics[width=1\linewidth]{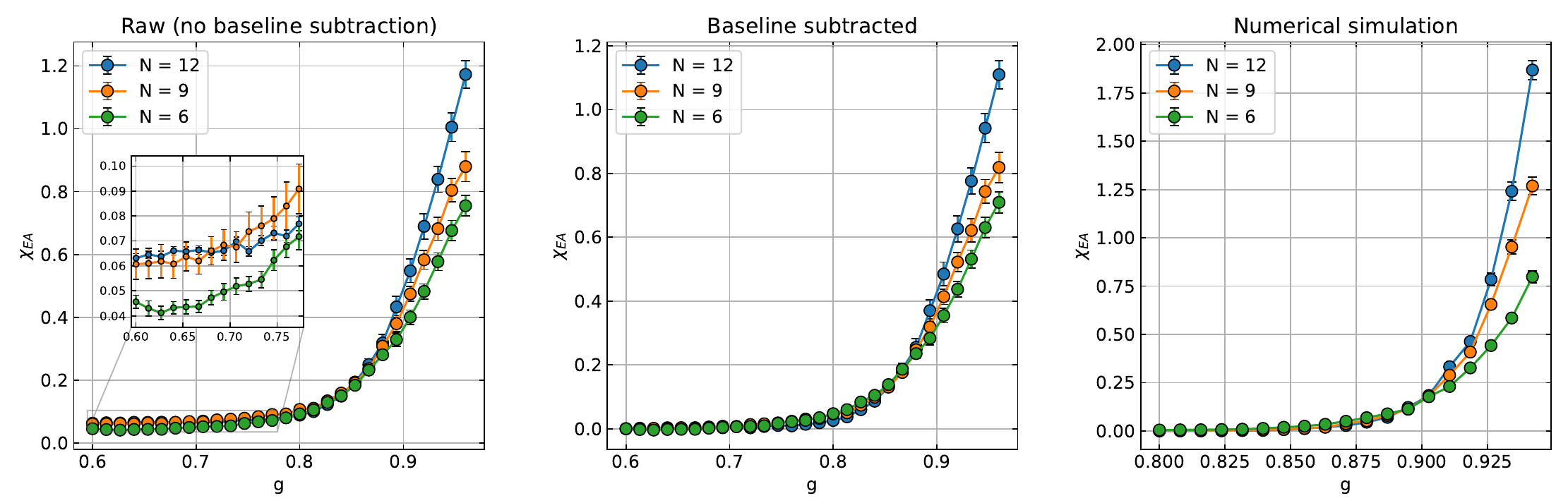}
    \caption{\textbf{Raw and baseline subtracted $\chi_{EA}$ data from main text and numerical simulation of our system.} To account for experimental noise that we attribute to bias due to shot noise, SPAM errors, flux transients from our tunable coupler pulses, and crosstalk, we performed a uniform baseline subtraction on each $\chi_{\text{EA}}$ curve by subtracting the value of $\chi_{\text{EA}}$ at the end point, $g=0.6$, deep in the thermal regime, at each $N$. We also compare these experimental results to noiseless numerical simulations of our system with 100 disorder averages also averaged from $t=60-70$ and with 20 initial states.}
    \label{fig:supp-chi}
\end{figure}

In Fig.~4 in the main body, we presented results for the finite system size analysis of our driven CCM. In particular, we considered the generalized spin-glass or Edwards-Anderson order parameter $\chi_{\text{EA}} = \frac{1}{N}\sum_{i \neq  j} |\expval{\hat{Z}_i \hat{Z}_j^\dag}|^2$ which measures the glassiness from spin correlations across our chain. In particular, we performed driven CCM experiments and measured this order parameter for chains of length $N=6,9,12$ across a range of kicking strengths $g$, for 5 Floquet cycles between $t=5$ and $t=10$, and over 13 initial states in each case. To try to ensure similar noise and couplings, the experiments were all performed across a two day period on the same set of qutrits and with similar coupling parameters chosen. 

In the main text, to account for experimental noise that we attribute to bias due to shot noise, SPAM errors, flux transients from our tunable coupler pulses, and crosstalk, we performed a uniform baseline subtraction on each $\chi_{\text{EA}}$ curve. As outlined in Fig.~\ref{fig:supp-chi}, this corresponded to uniformly subtracting the $\chi_{\text{EA}}$ value furthest into the thermal regime at $g=0.6$ from our data for each chain length. We note that this background subtraction does not change the inter-curve scaling, but was critical in helping identify an approximate experimental crossing for our order parameter. Additionally, it was motivated by our numerics (Fig.~\ref{fig:supp-chi}) which show that in the thermal regime, the $\chi_{\text{EA}}$ differences between chain lengths $N$ should be well below the noise floor of our experiment. We note that our experimental measurement of this order parameter can provide a useful estimate of the critical kicking strength of our phase transition $g_c$, but rigorous identification of this value, especially in the thermodynamic limit $N \rightarrow \infty$ remains the subject of future work.

\newpage
\newpage
\section{Theoretical and Numerical Results: Qutrits and the Driven CCM}
We begin by defining the qutrit Hilbert space which is spanned by $\ket{s}\in\{\ket{0},\ket{1},\ket{2}\}$ states of the transmon. The clock and shift operators are  
\begin{align}
\begin{split}
        \s\ket{s} &= \omega_3^s \ket{s} \;, \\
    \t \ket{s} &= \ket{(s+1)\mod 3} \;,
    \end{split}
\end{align}
where $\omega_3 = e^{2\pi i/3}$ is the cube root of unity. Their properties are summarized as follows
\begin{align}
    \s^3 &= \t^3 = I, & \s \t &= \omega_3 \t \s, & 1+ \omega_3 + \omega_3^2 = 0 \;.
\end{align}

\subsection{Mapping the cross-Kerr Hamiltonian to the CCM}\label{sec:theory-mapping}
In this section, we map the cross-Kerr interaction to the chiral clock model $U_{\text{CK}}=e^{i H_{\text{CK}}}\rightarrow e^{i H}$, where $H = H_{\text{int}} + H_{0}$, contains both the qutrit-qutrit interaction, $H_{\text{int}}$, and the single site, $H_0$, term. We perform baseband fast flux pulses on our tunable coupler to implement a cross-Kerr interaction between qutrits given by Eq.~\ref{eq:ck}. As our experiment works in a floquet implementation with gates rather than as an analog quantum simulation, to make our mapping between the experimentally realized unitary and the generators of our theoretical CCM, we define angles $\theta_{ij} = (2\pi \alpha_{ij} \times t) \in (-\pi,\pi]$. We can express the cross-Kerr Hamiltonian in the spin basis

\begin{align}
    H_{\text{CK}} = \theta_{11}\ket{11}\bra{11} + \theta_{12}\ket{12}\bra{12}+\theta_{21} \ket{21}\bra{21}+\theta_{22} \ket{22}\bra{22} = \sum_{m,n=0}^2 c_{mn} \s^m \otimes \s^n \;,
\end{align}
where the projections are
\begin{align}
    c_{mn} &= \frac{1}{9} Tr[(\s^m \otimes \s^n)^\dag H] = \frac{1}{9}\sum_{ij} \omega_3^{-im - jn} \bra{ij}H_{\text{CK}}\ket{ij} \\
    &= \frac{1}{9} (\theta_{11} \omega_3^{-m-n}+\theta_{12} \omega_3^{-m-2n}+ \theta_{21}\omega_3^{-2m-n} + \theta_{22} \omega_3^{-2m-2n}) \;.
\end{align}
Writing all independent terms 
\begin{align}
    c_{00} &=\frac{1}{9}(\theta_{11} +\theta_{12}+\theta_{21}+\theta_{22})\;, \\
    c_{01} &=\frac{1}{9}(\omega_3^2 (\theta_{11} +\theta_{21})+ \omega_3 (\theta_{12}+\theta_{22}))\;, \\
    c_{10} &= \frac{1}{9}(\omega^2(\theta_{11} +\theta_{12})+\omega_3(\theta_{21}+\theta_{22}))\;, \\
    c_{11} &=\frac{1}{9} (\omega_3 \theta_{11}  + \theta_{12} +\theta_{21} + \omega_3^2 \theta_{22})\;, \\
    c_{12} &=\frac{1}{9} (\theta_{11} +\theta_{12}\omega_3 + \theta_{21}\omega_3^2 + \theta_{22}) \;,
\end{align}
and comparing with the theory ansatz in the main text
\begin{align}
    H = J e^{i\theta} \hat Z_1 \otimes \hat Z_2^\dag + J^{\prime} e^{i\theta^\prime} \hat Z_1 \otimes \hat Z_2 + h_1 e^{i\varphi_1} \hat Z_1 \otimes I + h_2 e^{i\varphi_2} I \otimes \hat Z_2 + \text{h.c.}  + h_II \otimes I \;,
\end{align}
we obtain the relevant coefficients
\begin{align}
    J &= |c_{12}| \;, & \theta &= \arg({c_{12}})\;,  \\
    J^\prime &= |c_{11}| \;,& \theta^\prime &= \arg(c_{11})\;, \\
    h_1 &= |c_{10}|\;, & \varphi_1 &= \arg(c_{10})\;, \\
    h_2 &=|c_{01}| \;,& \varphi_2 &= \arg(c_{01}) \;,
\end{align}
and $h_I = c_{00}$ which only leads to a global phase.

\subsection{Time crystalline order in generalized clock model}

\subsubsection{Exactly Solvable Limit}
To investigate the properties of the driven chiral clock model, it is informative to first consider the exactly solvable limit at $g=1$. The Floquet unitary is $\mathcal U_F = e^{-iH} \bar{X}$, with
\begin{align}
     H &= \sum_i \underbrace{J_ie^{i\theta_i}\hat{Z}_i \hat{Z}^\dag_{i+1}}_{\mathbb{Z}_3  \text{ preserving}}+ { \sum_i J_i^\prime e^{i\theta_i^\prime} \hat{Z}_i \hat{Z}_{i+1} + h_i e^{i\varphi_i} \hat{Z}_i} + \text{h.c.} \;,
\end{align}
containing the combined clock spin-spin interaction $H_{\text{int}}$ and the single site $H_0$ term in the main text. We represent the spin basis of the entire chain by $\ket{ s} = \ket{s_1,s_2, \dots, s_L}$ where $s_i \in \{0,1,2\}$ is the spin at site $i$. The spin basis diagonalizes $H$ with the energy naturally written as the sum of two parts: $H \ket{ s} = (\varepsilon(s)+\varepsilon^\prime(s)) \ket{ s}$ where
\begin{align}
    \varepsilon(s) &= 2 \sum_i J_i \cos\bigg(\frac{2\pi}{3}(s_i-s_{i+1}) + \theta_i\bigg)\;, & \varepsilon^\prime(s) &= 2 \sum_i \bigg[J_i^\prime \cos\bigg( \frac{2\pi}{3} (s_i+s_{i+1})+ \theta_i^\prime\bigg) + h_i \cos(\frac{2\pi}{3}s_i + \varphi_i)\bigg] \;.
\end{align}
The first term $\varepsilon(s)$ only depends on the differences of spins and is therefore invariant under a global kick $\bar X = \prod_i X_i$. This term arises out of the $\mathbb{Z}_3$ symmetry preserving part of $H$: $H_{\mathbb{Z}_3}= \sum_i J_ie^{i\theta_i}\hat{Z}_i \hat{Z}^\dag_{i+1} + \text{h.c.}$, satisfying $[H_{\mathbb{Z}_3},\bar{X}]= [H_{\mathbb{Z}_3},\bar{X}^2]=0$. A consequence is that $H_{\mathbb{Z}_3}$ has the same energy $\varepsilon( s)$ for $\{ \ket{ s}, \bar{X}\ket{ s}, \bar{X}^2 \ket{ s}\}=\{ \ket{ s}, \ket{ { s + 1}}, \ket{ {s + 2}}\}$. The Floquet unitary in this basis is
\begin{align}
    \mathcal U_F = e^{-i\varepsilon({ s})}\begin{bmatrix}
      0 & 0 & e^{-i\varepsilon^\prime(s)}  \\
      e^{-i\varepsilon^\prime(s+1)} & 0 & 0 \\
      0 & e^{-i\varepsilon^\prime(s+2)} &0
    \end{bmatrix} \;.
\end{align}
We diagonalize it to obtain the corresponding quasienergies and vectors $\mathcal U_F\ket{\varepsilon_i}=e^{-i \varepsilon_i} \ket{\varepsilon_i}$ where
\begin{align}
    \varepsilon_0(s) &= \varepsilon(s), & \ket{\varepsilon_0(s)} &= \frac{1}{\sqrt{3}} (e^{-i\varepsilon^\prime(s)}\ket{s}+ e^{i\varepsilon^\prime(s+2)}\ket{s+1} + \ket{s+2}) \;, \\
    \varepsilon_1(s) &= \varepsilon(s) + 2\pi/3, & \ket{\varepsilon_1(s)} &= \frac{1}{\sqrt{3}} (\omega_3 e^{-i\varepsilon^\prime(s)}\ket{s} + \omega_3^2 e^{i\varepsilon^\prime(s+2)} \ket{s+1} + \ket{s+2}) \;, \\
    \epsilon_2(s) &= \varepsilon(s) - 2\pi/3, & \ket{\varepsilon_2(s)} &= \frac{1}{\sqrt{3}} (\omega_3^2 e^{-i \varepsilon^\prime(s)}\ket{s} + \omega_3 e^{i \varepsilon^\prime(s+2)} \ket{s+1}+ \ket{s+2}) \;.
\end{align}
For simplification, we used an additional constraint: $\varepsilon^\prime(s) + \varepsilon^\prime(s+1)+\varepsilon^\prime(s+2) = 0$, following from $H + \bar{X} H \bar{X}^\dag + \bar{X}^2 H \bar{X}^{\dag 2} = 3 H_{\mathbb{Z}_3}$ which we explicitly prove below. We see that the eigenstates are long range ordered cat-like states. The  quasienergies form triplets separated by $2\pi/3$ and depend only on the $\mathbb{Z}_3$ symmetry preserving part of the Hamiltonian. 
Our results for the general chiral clock model are consistent with those from  Ref.~\cite{surace2019floquet} that consider a similar model.
\begin{figure}
    \centering
    \includegraphics[width=1\columnwidth]{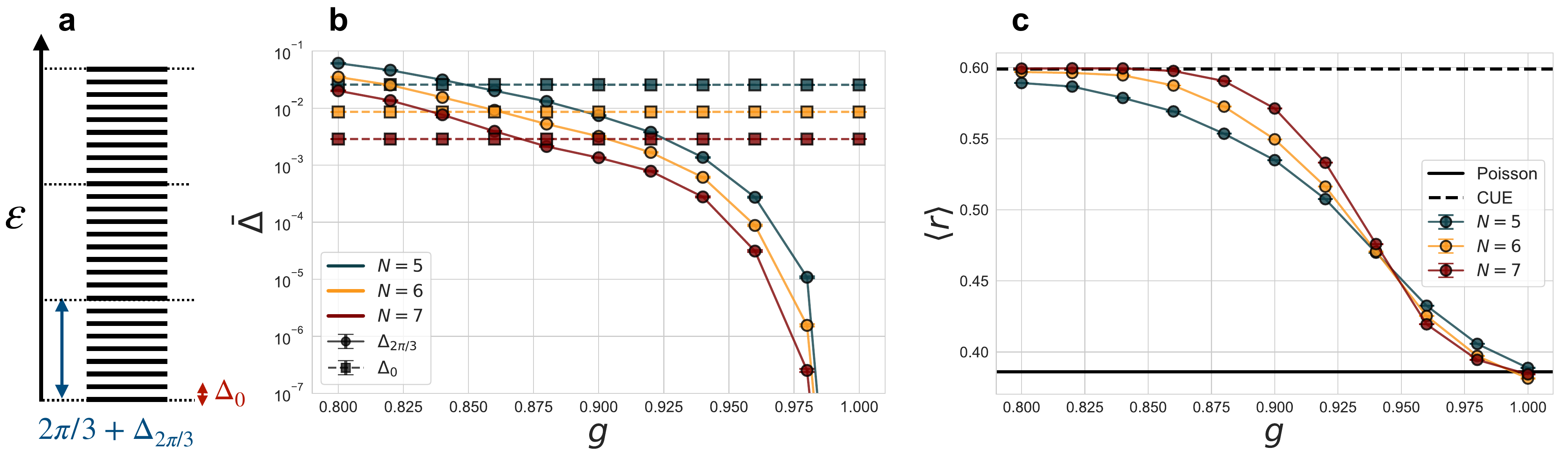}
    \caption{(a) Spectrum and definition of quasienergy gaps. (b) The average spectral gap between adjacent quasienergies $\Delta_0$ and the error in triplet pairing $\Delta_{2\pi/3}$ as a function of kick parameter $g$. (c) Average statistics $\langle r \rangle$ over quasienergies and disorder realizations as a function of kick $g$. The Hamiltonian parameters are chosen to be $J, J^\prime, h \in [0.1, 0.3]$ and random $\theta, \theta^\prime, \varphi$ averaged over 1000 disorder instances.}
    \label{fig:pairing}
\end{figure}
\textit{Dynamical decoupling of $\mathbb{Z}_3$ breaking terms ---} After three Floquet cycles we have the unitary
\begin{align}
\begin{split}
     \mathcal U_F^3 &= e^{-iH} \bar{X} e^{-iH} \bar{X} e^{-iH} \bar{X}  = e^{-i (H + \bar{X}H\bar{X}^\dag + \bar{X}^2 H \bar{X}^{\dag 2})} \;,
    \end{split} 
\end{align}
where we used $\bar{X}^3 = I$ and $H$ is diagonal. Notice that the $\mathbb Z_3$ breaking terms rotate by integer multiples of $\omega_3$: $\bar{X} \hat{Z}_i \bar{X}^\dag = \omega_3^{2} \hat{Z}_i$, $\bar{X} \hat{Z}_i \hat{Z}_{i+1} \bar{X}^\dag = \omega_3 \hat{Z}_i \hat{Z}_{i+1}$. Together with the fact that $[H_{\mathbb{Z}_3}, \bar X] = 0$, we have 
\begin{align}
    H + \bar{X} H \bar{X}^\dag + \bar{X}^2 H \bar{X}^{\dag 2} &= 3 H_{\mathbb{Z}_3} + \bigg( (1+ \omega_3 + \omega_3^2)J_i^\prime e^{i\theta_i^\prime} \hat{Z}_i \hat{Z}_{i+1} + (1+\omega_3^2 + \omega_3)h_i e^{i\phi_i^\prime} \hat{Z}_i + \text{h.c.} \bigg) \\
    &= 3 H_{\mathbb{Z}_3} \;.
\end{align}
Therefore, near $g \sim 1$, the $\mathbb{Z}_3$ breaking terms dynamically decouple out, leaving the quasienergies to be dictated by the $\mathbb{Z}_3$ preserving part $ \mathcal U_F^3 = e^{-3iH_{\mathbb{Z}_3}}$. This is similar to the Ising case when $\mathbb{Z}_2$ breaking terms decouple out~\cite{mi2022time, PhysRevB.94.085112, ippoliti2021many}.
\newpage 

\subsubsection{Eigenstate Order and Many Body Localization}

The eigenstate order must be stable beyond the trivial $g=1$ point for true time crystalline behavior. 
To see this, we numerically diagonalize the Floquet unitary $\mathcal U_F \ket{\varepsilon_i}=e^{i\varepsilon_i}\ket{\varepsilon_i}$ as a function of the kick parameter $g$. For each quasienergy $\varepsilon_i \in [-\pi, \pi)$, we denote
\begin{align}
    \Delta^i_0 &= |\varepsilon_{i+1}-\varepsilon_i| \;, & \Delta^i_{2\pi/3} &= |\varepsilon_{i}-\varepsilon_{i+\mathcal N/3} - 2\pi/3| \;,
\end{align}
where $\mathcal N = 3^N$ the dimension of the Hilbert space. Here $\Delta^i_0$ quantifies the gap in adjacent quasienergies and $\Delta^i_{2\pi/3}$ represents the error in triplet pairing as depicted in Fig.~\ref{fig:pairing}(a). Since eigenstate order is necessary for time crystalline behavior, the error in triplet pairings must be small as compared to the gap between subsequent quasienergies. We average over quasienergies and disorder configurations and respectively show $\bar \Delta_0, \bar \Delta_{2\pi/3}$ as a function of the kicking strength $g$ in Fig.~\ref{fig:pairing}(b). From these, we see that the eigenstate order persists beyond $g=1$ which confirms the time-crystalline behavior observed in the experiment. While $\bar \Delta_0$ is independent of $g$ and only depends on system size $N$, $\bar \Delta_{2\pi/3}$ shows $g$-dependent behavior. At $g=1$ the error in triplet pairing approaches machine precision, confirming the analytical derivation above and the eigenstate pairing. Furthermore, the triplet pairing error becomes larger than the quasienergy gap around $g\sim 0.86$, which is consistent with the experimental observation in Fig.~2 in the main text, indicated by the Fourier transform losing its sharp peak.

Next, we study ergodicity breaking due to disorder~\cite{ponte, pal}. We investigate energy level statistics of the quasienergy gap $\Delta_0^i$ by calculating ratio \begin{align}
    r = \frac{\text{min}(\Delta_0^i,\Delta_0^{i+1})}{\text{max}(\Delta_0^i,\Delta_0^{i+1})} \;.
\end{align}
We plot the average $\langle r \rangle $ over quasienergies and disorder configurations in Fig.~\ref{fig:pairing}(c) for various chain lengths. Near $g \sim 1$, as the system size increases, the ratio appears to converge to Poisson statistics value $\langle  r\rangle_{\text{Poisson}}\approx 0.386$ indicating ergodicity breaking and the system being localized in the MBL phase. On the other extreme at low $g$, the system is ergodic or thermal as indicated by its convergence to the circular unitary ensemble (CUE) where $\langle r \rangle_{\text{CUE}}\approx 0.599$. For the typical disorder range in the experiment, the crossover to the MBL region happens after $g \sim 0.93$. This is consistent with the eigenstate pairing error in panel (b) which rapidly decreases beyond $g\sim 0.93$ indicating stable period tripling for long times.
\newpage
\subsubsection{Numerical simulation of the Time Crystal}
\begin{figure}[h!]
    \centering
    \includegraphics[width=1\columnwidth]{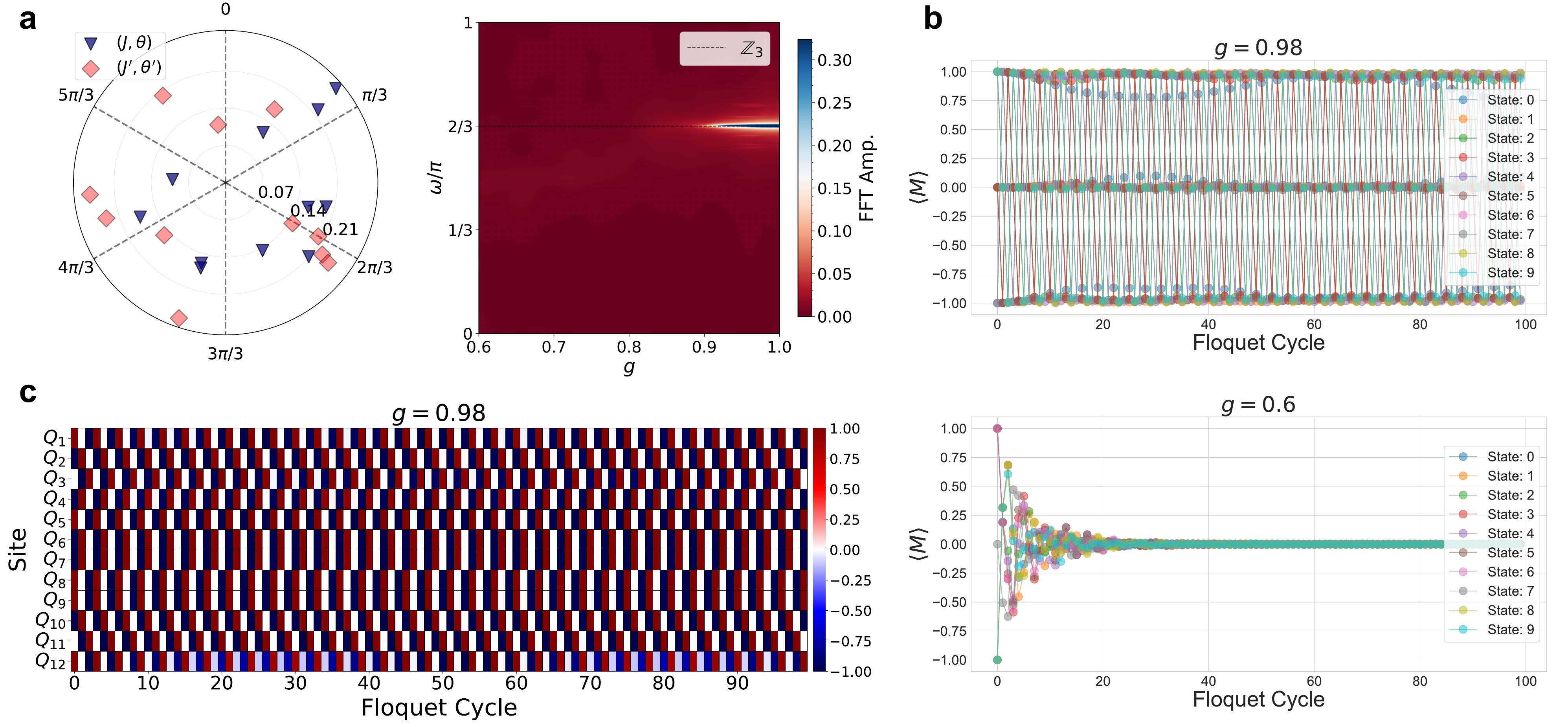}
    \caption{\textbf{Numerical simulation of driven CCM Floquet system}. (a) Coupling parameters for a chain of $12$ clock spins. In addition, the onsite field parameters are $h \in [0.1,0.3]$ and randomly chosen $\varphi \in (-\pi, \pi]$. FFT response of magnetization $\langle M \rangle$ for the bulk qutrit $Q_6$. (b) Magnetization response of the bulk qutrit $Q_6$ for random initial product states in MBL DTC ($g=0.98$) and thermal regime ($g=0.6$). (c) Bulk response of the entire chain at $g=0.98$.}
    \label{fig:TCnumerics}
\end{figure}

Armed with this theoretical understanding, we now present a numerical simulation of the driven chiral clock model for a chain of $12$ qutrits using exact tensor network simulation. Figure~\ref{fig:TCnumerics}(a) depicts the random coupling parameters chosen within the experimentally accessible range. We show the Fourier transform of magnetization $\langle M \rangle$ response for the bulk site $Q_6$ as a function of $g$. We confirm that there is a sharp response exactly at $\omega =2\pi/3$ that shows robust period tripling and time crystalline behavior as discussed in the main text. We show that the time crystalline phase remains largely robust to the choice of initial states by considering 10 random initial trit string states in Fig.~\ref{fig:TCnumerics}(b). Furthermore, in the thermal regime, the oscillations die down within the time scales observed in the experiment and remain close to zero. Finally, we show the response of the entire chain in Fig.~\ref{fig:TCnumerics}(c) highlighting the collective response of the bulk. 
\newpage
\subsubsection{Chirality controlled stability of the CCM DTC}
\begin{figure}[h!]
    \centering
    \includegraphics[width=1\columnwidth]{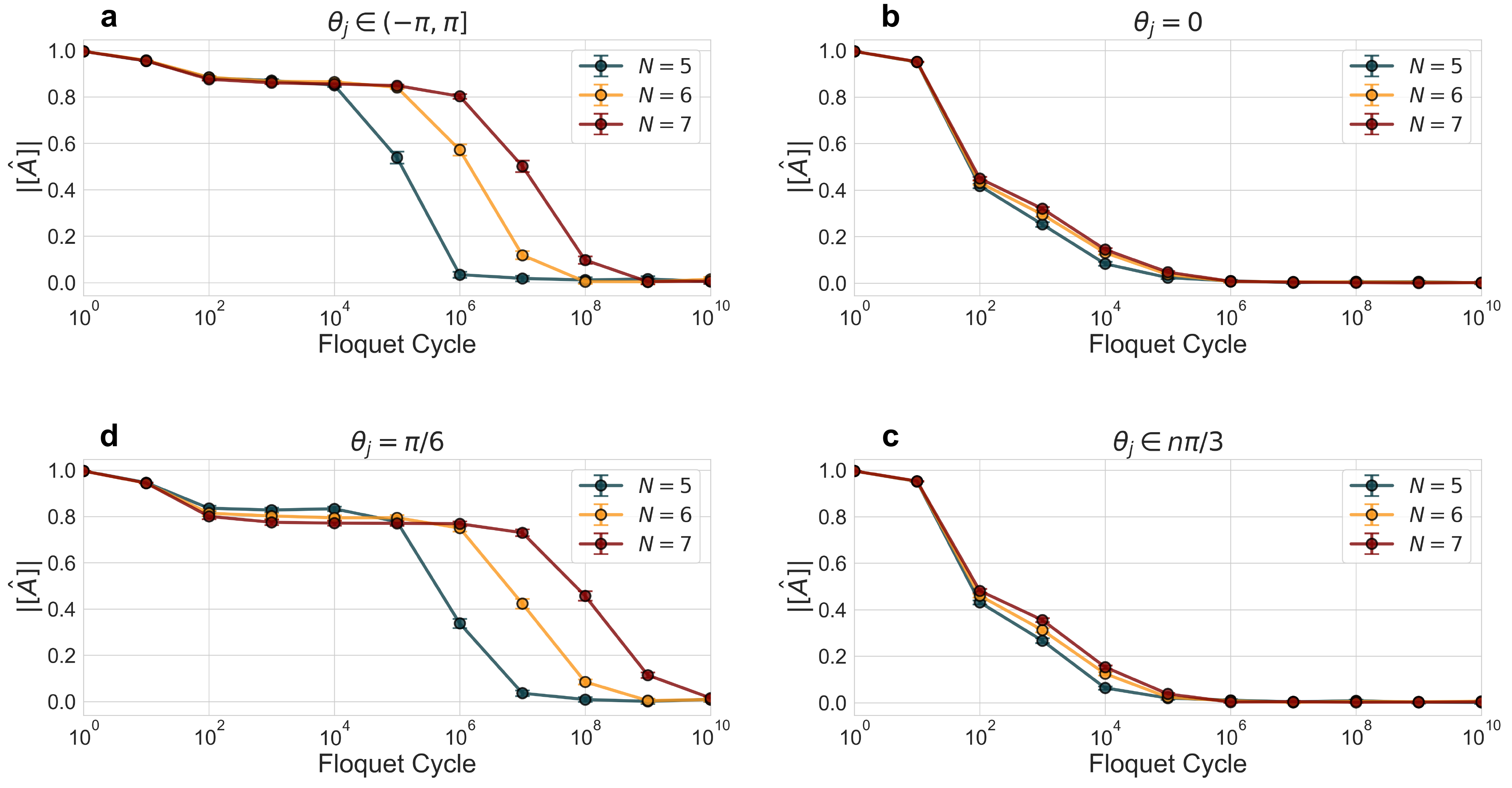}
    \caption{\textbf{Chirality controlled stability and system size scaling.} Using exact diagonalization, we calculate the autocorrelator $\hat{A}(t)=\omega_3^{-t} \langle Z(t) Z^\dagger(0)\rangle$ as a function of time for varying chain lengths. We plot the average $|[\hat{A}]|$ across sites, $100$ disorder configurations and $20$ random intial trit string states. For each case, we set $g=0.98$, $(J_j,J^\prime_j,h_j) \in [0.1,0.3]$ and $(\theta^\prime_j,\varphi_j) \in (-\pi,\pi]$, with different chiral angle (a) $\theta_{j} \in (-\pi,\pi]$, (b) $\theta_j = 0$, (c) $\theta_j \in n \pi/3$, where $n=0,\dots, 6$ and (d) $\theta_j = \pi/6$.}
    \label{fig:chiral_auto}
\end{figure}

We now highlight the special role of chirality in determining the stability of the $\mathbb{Z}_3$ DTC. As argued in the main text, distinct spin-domain configurations become degenerate at $\theta=n\pi/3$, where $n$ is an integer, leading to degeneracies in the quasienergy spectrum. Consequently, even a small perturbation can hybridize the corresponding long-range cat-like eigenstates, giving rise to short-range correlated states. Since a stable $\mathbb{Z}_3$ DTC must exhibit robust period tripling for all short-range (physical) states, it is necessary that no eigenstate of the Floquet unitary be short-range correlated. The emergence of such short-range correlated eigenstates therefore destabilizes the $\mathbb{Z}_3$ DTC in general.

We show this chirality dependent behavior by calculating the autocorrelator  $\hat{A}(t)=\omega_3^{-t} \langle Z(t) Z^\dagger(0)\rangle$, averaged over random product state initializations for a range of system sizes in Fig.~\ref{fig:chiral_auto}. For the kick strength $g=0.98$, we plot the averaged autocorrelator across sites, $100$ disorder configurations and $20$ random initial trit string states for different chiral angle configurations. We confirm that the period tripling stability increases with system size for randomly chosen $\theta$ that avoid $n\pi/3$ points, indicating stable DTC phase in Fig.~\ref{fig:chiral_auto}(a). In contrast, when $\theta=0$ or randomly chosen at $n\pi/3$ points, the autocorrelator quickly decays in Fig.~\ref{fig:chiral_auto}(b,c), with no significant system size dependence. In panel (d), we choose maximal chiral interaction at $\theta=\pi/6$ which maximizes the asymmetry in the domain wall energies. Here, robust period-tripling persists and strengthens with increasing chain length. Notably, this regime is even more stable than the disordered-$\theta$ case shown in panel (a). These results demonstrate that the stability of the DTC in the CCM is governed primarily by chirality through its control of domain-wall energetics, rather than by disorder alone.

\begin{figure}[h!]
    \centering
    \includegraphics[width=1\columnwidth]{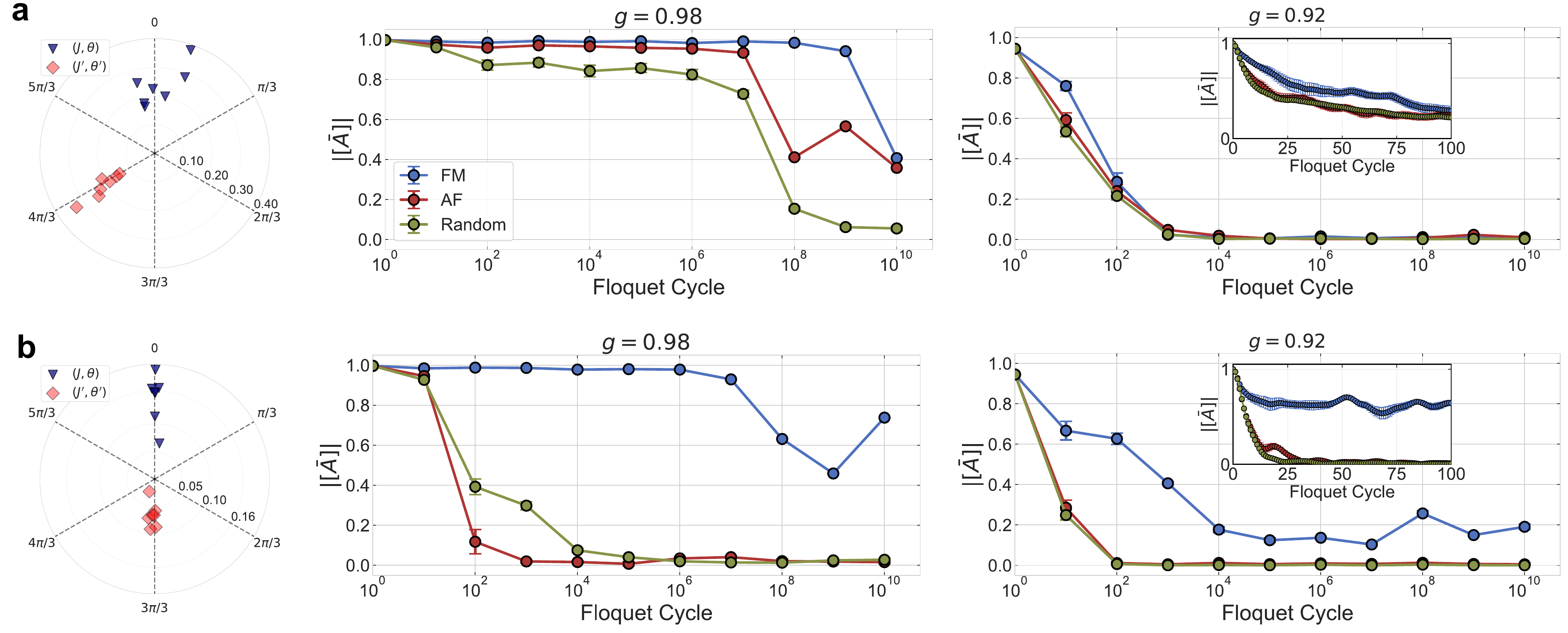}
    \caption{\textbf{Chirality induced stability and state dependence.} We calculate the autocorrelator $\hat A(t)$ using exact diagonalization for the $8$ qutrit experiment in Fig.~3 in the main text. We plot the average $|[\hat A]|$ over sites and FM, AF and random initial trit string states, comparing the cases when chiral angle is (a) non-zero and disordered (see left plot for values), to that when it is set to (b) zero.}
    \label{fig:chiral_ex}
\end{figure}

We now simulate the experimental protocol discussed in the main text, which highlights chirality-controlled domain-wall physics through state-dependent stability in Fig.~\ref{fig:chiral_ex}. We compare the driven CCM with chiral interactions ($\theta\neq 0$) in panel (a) to the non-chiral case in panel (b), where dynamical decoupling pulses are used to set $\theta=0$. At $\theta = 0$, the quasienergy spectrum recovers time-reversal and spatial inversion symmetry. As discussed in the main text, the two types of antiferromagnetic (AF) domain walls, (01,12,20) and (10,21,02), become degenerate, while ferromagnetic (FM) configurations (00,11,22) remain energetically distinct. Consequently, across the full qutrit chain, FM-like cat states remain spectrally isolated, whereas cat states associated with random or AF configurations hybridize. We point out that FM states are not special, tuning $\theta=\pm\pi/3$, instead will result in the corresponding AF state to remain energetically distinct and cat-states with configurations containing the other two domain walls (or lack thereof) will hybridize.

For the chiral case ($\theta\neq 0$), Fig.~\ref{fig:chiral_ex}(a) shows that the autocorrelator exhibits qualitatively similar behavior across all initial trit-string states. In particular, at $g=0.98$, robust period-tripling persists over long times. In contrast, in the non-chiral case shown in panel (b), stable period tripling is observed only for FM initial states, while other states decay rapidly, indicating instability. To facilitate experimental observation, we also consider a weaker drive (g=0.92), where the distinction between the two cases becomes more pronounced at shorter times.

\newpage
\putbib[Z3Driven]
\end{bibunit}

\end{document}